\documentclass[a4paper]{aa}
\usepackage{aabib, epsfig}

\newcommand{\ignore}[1]{}   

\def\ddeg{\hbox{.\hskip-3pt $^\circ$}}

\begin{document}
\thesaurus{
	03.09.3; 	
	03.19.3;	
	05.01.1;	
	10.07.1;	
	03:20.4;	
	03:20.7		
}	
	
\title{GAIA: Composition, Formation and Evolution of the Galaxy}

\author{M.A.C.~Perryman\inst{1}
\and K.S.~de Boer\inst{2}
\and G.~Gilmore\inst{3}
\and E.~H{\o}g\inst{4}
\and M.G.~Lattanzi\inst{5}
\and L.~Lindegren\inst{6}
\and X.~Luri\inst{7}
\and F.~Mignard\inst{8}
\and O.~Pace\inst{9}
\and P.T.~de Zeeuw\inst{10}
}

\institute{
Astrophysics Division, 
Space Science Department of ESA, ESTEC,
Postbus 299, NL-2200 AG Noordwijk, 
The Netherlands
\and 
%
Sternwarte Univ Bonn,
Auf dem Hugel 71,
Bonn 53121,
Germany
\and
%
University of Cambridge,
Institute of Astronomy,
Madingley Road,
Cambridge CB3 0HA,
United Kingdom
\and
%
Copenhagen University Observatory,
Juliane Maries Vej 30,
Copenhagen OE,
DK-2100, Denmark
\and
%
Osservatorio Astronomico di Torino,
Strada Osservatorio 20,
Pino Torinese (TO),
I-10025 Italy
\and
%
Lund Observatory,
Box 43,
Lund 22100,
Sweden
\and
%
Universitat de Barcelona,
Departament d'Astronomia i Meteorologia,
Avda Diagonal 647,
Barcelona 08028,
Spain
\and
%
Observatoire de la C\^ote d'Azur,
CERGA,
Avenue Copernic,
Grasse 06130, 
France
\and
%
Future Projects Division of ESA, ESTEC,
Postbus 299, NL-2200 AG Noordwijk, 
The Netherlands
\and 
%
Sterrewacht,
Jan Hendrik Oort Building,
Postbus 9513,
Leiden 2300 RA, 
The Netherlands
}


\date{Received date ; accepted date }

\titlerunning{GAIA: Formation of the Galaxy}
\authorrunning{M.A.C.$\,$Perryman et al.}

\maketitle 
\begin{abstract}
The GAIA astrometric mission has recently been approved as one of the
next two `cornerstones' of ESA's science programme, with a launch date
target of not later than mid-2012. GAIA will provide positional and
radial velocity measurements with the accuracies needed to produce a
stereoscopic and kinematic census of about one billion stars
throughout our Galaxy (and into the Local Group), amounting to about
1~per cent of the Galactic stellar population.  GAIA's main scientific
goal is to clarify the origin and history of our Galaxy, from a
quantitative census of the stellar populations. It will advance
questions such as when the stars in our Galaxy formed, when and how it
was assembled, and its distribution of dark matter. The survey aims
for completeness to $V=20$~mag, with accuracies of about 10~$\mu$as at
15~mag. Combined with astrophysical information for each star,
provided by on-board multi-colour photometry and (limited)
spectroscopy, these data will have the precision necessary to quantify
the early formation, and subsequent dynamical, chemical and star
formation evolution of our Galaxy. Additional products include
detection and orbital classification of tens of thousands of
extra-Solar planetary systems, and a comprehensive survey of some
$10^5-10^6$ minor bodies in our Solar System, through galaxies in the
nearby Universe, to some 500\,000 distant quasars. It will provide a
number of stringent new tests of general relativity and cosmology. The
complete satellite system was evaluated as part of a detailed
technology study, including a detailed payload design, corresponding
accuracy assesments, and results from a prototype data reduction
development.


\end{abstract}

\section{Introduction}
\label{sec:intro}

Understanding the details of the Galaxy in which we live is one of the
great intellectual challenges embraced by modern science. Our Galaxy
contains a complex mix of stars, planets, interstellar gas and dust,
radiation, and the ubiquitous dark matter. These components are widely
distributed in age (reflecting their birth rate), in space (reflecting
their birth places and subsequent motions), on orbits (determined by
the gravitational force generated by their own mass), and with complex
distributions of chemical element abundances (determined by the past
history of star formation and gas accretion).

Astrophysics has now developed the tools to measure these
distributions in space, kinematics, and chemical abundance, and to
interpret the distribution functions to map, and to understand, the
formation, structure, evolution, and future of our entire Galaxy. This
potential understanding is also of profound significance for
quantitative studies of the high-redshift Universe: a well-studied
nearby template galaxy would underpin the analysis of unresolved
galaxies with other facilities, and at other wavelengths.

Understanding the structure and evolution of the Galaxy requires three
complementary observational approaches: (i) a census of the contents
of a large, representative, part of the Galaxy; (ii) quantification of
the present spatial structure, from distances; (iii) knowledge of the
three-dimensional space motions, to determine the gravitational field
and the stellar orbits. Astrometric measurements uniquely provide 
model-independent distances and transverse kinematics, and form the basis
of the cosmic distance scale.  Complementary radial velocity and 
photometric information are required to complete the kinematic and 
astrophysical information about the individual objects observed.

Photometry, with appropriate astrometric and astrophysical
calibration, gives a knowledge of extinction, and hence, combined with
astrometry, provides intrinsic luminosities, spatial distribution
functions, and stellar chemical abundance and age information. Radial
velocities complete the kinematic triad, allowing determination of
dynamical motions, gravitational forces, and the distribution of 
invisible mass. The GAIA mission will provide all this information.
  
Even before the end of the Hipparcos mission, a proposal for an
ambitious follow-on space astrometry experiment was submitted to ESA
(Roemer: \cite{hog93}; \cite{lbg+93a}; \cite{hl94}). The idea of using
CCDs as a modulation detector behind a grid (\cite{hl93}), similar to 
Hipparcos, was replaced by the more powerful option adopted for 
Roemer (\cite{hog93}) where CCDs measure the direct stellar images in 
time-delayed integration (TDI) mode in the scanning satellite. A more
ambitious interferometric mission, GAIA, was proposed and subsequently
recommended as a cornerstone mission of the ESA science programme by
the Horizon 2000+ Survey Committee in 1994 (\cite{bat94}).  The GAIA
proposal demonstrated that accuracies of 10~$\mu$as at 15~mag were
achievable using a small interferometer (\cite{lpb+93b}; \cite{lp96}).

The European scientific community and ESA have now completed a
detailed study of the science case and instrument design,
identifying a number of further improvements, including reverting to 
full-aperture telescopes (\cite{hog95a};
\cite{hog95b}). The results demonstrate that unique and fundamental
advances in astrophysics are technically achievable on the proposed
time-scales, and within a budget profile consistent with the current
ESA cornerstone mission financial envelope.

GAIA will be a continuously scanning spacecraft, accurately measuring
one-dimensional coordinates along great circles in two simultaneous
fields of view, separated by a well-known angle. The payload utilises
a large but feasible CCD focal plane assembly,  passive thermal
control, natural short-term instrument stability due to the Sun shield
and the selected orbit, and a robust payload design. The telescopes
are of moderate size, with no specific manufacturing complexity. The
system fits within a dual-launch Ariane~5 configuration, without
deployment of any payload elements. The study identifies a `Lissajous'
orbit at L2 as the preferred operational orbit, from where about
1~Mbit of data per second is returned to the single ground station
throughout the 5-year mission. A comprehensive accuracy assessment has
validated the proposed payload and the subsequent data reduction. 

This paper provides a summary of the key features of the improved GAIA
design, and the resulting scientific case, evaluated during the recent
study phase (\cite{esa-2000-4}; see also http://astro.estec.esa.nl/GAIA).
A comparison between the scientific goals of GAIA, and other post-Hipparcos
space astrometry missions, is given in Section~\ref{sec:other-missions}.

\section{Scientific Goals}
\label{sec:scientific-goals}

\subsection{Structure and Dynamics of the Galaxy}

The primary objective of the GAIA mission is to observe the physical 
characteristics, kinematics and distribution of stars over a large 
fraction of the volume of our Galaxy, with the goal of achieving a full
understanding of its dynamics and structure, and consequently
its formation and history (see, e.g., \cite{gwk89}; \cite{maj93}; 
\cite{iwg+97}; \cite{wgf97}; \cite{zee99}; as well as 
extensive details of the scientific case given in \cite{esa-2000-4}). An
overview of the main Galaxy components and sub-populations is given in
Table~\ref{tab:tracers}, together with requirements on astrometric
accuracy and limiting magnitude.

\begin{table*}[tbh]
\caption{Some Galactic kinematic tracers, with corresponding 
limiting magnitudes and required astrometric accuracy. For various 
tracers (Column~1), Columns~2--6 indicate relevant values of 
parameters leading to the typical range of $V$~magnitudes over which 
the populations must be sampled (Columns~7--8). These results
demonstrate that the faint magnitude limit of GAIA is essential for probing
the different Galaxy populations, while the astrometry accuracies in
Columns~9--12 demonstrate that GAIA will meet the scientific goals
(based on \protect\cite{gh95}).}  
\vspace{5pt}
\label{tab:tracers}
\begin{center} 
\leavevmode
\scriptsize
\begin{tabular}{lrrrrrccrccc}
\hline 
&&&&&&&&& &&\\[-5pt]
 (1) & (2) & (3) & (4) & (5) & (6) & (7) & (8) & (9) & (10) & (11) &
(12) \\
 Tracer\qquad \qquad & 
	M$_{\rm V}$ & \hfil $\ell$ \hfil & \hfil $b$ \hfil & \hfil
$d$ \hfil & A$_{\rm V}$ & \hfil  V$_{1}$ \hfil & \hfil V$_{2}$ \hfil &
$\epsilon_{\rm T}$ \hfil &  
$\sigma_{\mu_1}$ & $\sigma^{\prime}_{\mu_1}$ &
$\sigma^{\prime}_{\pi_1}$ \\
&  mag  &  deg  &  deg  &  kpc  &  mag  &  mag  & 
         mag  &  km/s  &  $\mu$as/yr  & -- & --\\[4pt]
\hline 
&&&&&&&&& &&\\
\noalign{\vskip -5pt}

{Bulge:} &&&&&&&&& &&\\
\hspace{10pt} gM & --1 & 0 & $<20$ & 8 & 2--10 & 15 & 20 & 100 & 10&0.01 &0.10\\
\hspace{10pt} HB & +0.5 & 0 & $<20$ & 8 & 2--10 & 17 & 20 & 100 & 20&0.01 &0.20\\
\hspace{10pt} MS turnoff & +4.5 & 1 & $-4$ & 8 & 0--2 & 19 & 21 & 100 
&60&0.02 &0.6\phantom{0}\\
&&&&&&&&& &&\\[-5pt]
{Spiral arms:} &&&&&&&&& &&\\
\hspace{10pt} Cepheids & --4 & all & $<10$ & 10 & 3--7 & 14 & 18 & 7 & \phantom{0}5&0.03 &0.06\\
\hspace{10pt} B--M Supergiants & --5 &  all & $<10$& 10 & 3--7& 13 &17 &7& \phantom{0}4&0.03 &0.05\\
\hspace{10pt} {Perseus arm (B)}& 
          --2 & 140 & $<10$ & 2 & 2--6 & 12 & 16 & 10 & \phantom{0}3&0.01&0.01\\
&&&&&&&&& &&\\[-5pt]
{Thin disk: } &&&&&&&&& &&\\
\hspace{10pt} gK & --1 & 0 & $<15$ & 8 & 1--5 & 14 & 18 & 40 & \phantom{0}6&0.01 &0.06\\
\hspace{10pt} gK & --1 & 180 & $<15$ & 10 & 1--5 & 15 & 19 & 10 &  \phantom{0}8&0.04 &0.10\\
&&&&&&&&& &&\\[-5pt]
Disk warp (gM)& --1 & all & $<20$ & 10 &1--5 &15 &19 & 10 & \phantom{0}8& 0.04 &0.10\\
&&&&&&&&& &&\\[-5pt]
Disk asymmetry (gM)& --1 & all & $<20$ & 20 &1--5 &16 &20 & 10 
&15& 0.14 &0.4\phantom{0}\\
&&&&&&&&& &&\\[-5pt]
{Thick disk:} &&&&&&&&& &&\\
\hspace{10pt} Miras, gK & --1 & 0 & $<30$ & 8 & 2 & 15 & 19 & 50 
&10&0.01 &0.10\\
\hspace{10pt} HB & +0.5 & 0 & $<30$ & 8 & 2 & 15 & 19 & 50 & 20&0.02 &0.20\\
\hspace{10pt} Miras, gK & --1 & 180 & $<30$ & 20 & 2 & 15 & 21 & 30 
&25&0.08 &0.65\\
\hspace{10pt} HB & +0.5 & 180 & $<30$ & 20 & 2 & 15 & 19 & 30 & 60&0.20 
&1.5\phantom{0}\\
&&&&&&&&& &&\\[-5pt]
{Halo:} &&&&&&&&& &&\\
\hspace{10pt} gG & --1  & all & $<20$ & 8 & 2--3 & 13 & 21 & 100 & 10&0.01 &0.10\\
\hspace{10pt} HB & +0.5 & all & $>20$ & 30 & 0 & 13 & 21 & 100 &35&0.05& 
1.4\phantom{0}\\
&&&&&&&&& &&\\[-5pt]
{Gravity, $\cal {K}_Z$: } &&&&&&&&& &&\\
\hspace{10pt} dK& 
        +7--8 & all & all & 2 & 0 & 12 & 20 & 20 & 60&0.01 &0.16\\
\hspace{10pt} dF8-dG2& 
        +5--6 & all & all & 2 & 0 & 12 & 20 & 20 & 20&0.01 &0.05\\
&&&&&&&&& &&\\[-5pt]
Globular clusters (gK)& +1 & all & all & 50 & 0 & 12 & 21 & 100 & 10&0.01 &0.10\\
&&&&&&&&& &&\\[-5pt]
\hspace{10pt} internal kinematics (gK)& +1 & all & all & 8 & 0 & 13 & 17 & 15 & 10&0.02 &0.10\\
&&&&&&&&& &&\\[-5pt]
Satellite orbits (gM)& --1 & all & all & 100 & 0 & 13 & 20 & 100 & 60&
0.3\phantom{0}& 8\phantom{.00}\\
\hline 
\end{tabular}
\end{center}
\end{table*}

\subsection{The Star Formation History of our Galaxy}

A central element the GAIA mission is the determination of
the star formation histories, as described by the temporal evolution
of the star formation rate, and the cumulative numbers of stars
formed, of the bulge, inner disk, Solar neighbourhood, outer disk and
halo of our Galaxy (e.g.\ \cite{hgv99}). Given such information, together 
with the
kinematic information from GAIA, and complementary chemical abundance
information, again primarily from GAIA, the full evolutionary history
of the Galaxy is determinable (e.g.\ \cite{fre93}; \cite{gil99x}).

Determination of the relative rates of formation of the stellar 
populations in a large spiral, typical of those galaxies which 
dominate the luminosity in the Universe, will provide for the first 
time quantitative tests of galaxy formation models. 
Do large galaxies form from accumulation of many
smaller systems which have already initiated star formation? Does star
formation begin in a gravitational potential well in which much of the
gas is already accumulated? Does the bulge pre-date, postdate, or is
it contemporaneous with, the halo and inner disk? Is the thick disk a
mix of the early disk and a later major merger? Is there a radial age
gradient in the older stars? Is the history of star formation
relatively smooth, or highly episodic?  Answers to such questions will 
also provide a template for analysis of data on unresolved stellar 
systems, where similar data cannot be obtained.

\subsection{Stellar Astrophysics}

GAIA will provide distances of unprecedented accuracy for all types of
stars of all stellar populations, even those in the most rapid
evolutionary phases which are very sparsely represented in the Solar
neighbourhood. All parts of the Hertzsprung--Russell diagram will be
comprehensively calibrated, from pre-main sequence stars to white
dwarfs and all transient phases; all possible masses, from brown
dwarfs to the most massive O~stars; all types of variable stars; all
possible types of binary systems down to brown dwarf and planetary
systems; all standard distance indicators, etc. This extensive amount
of data of extreme accuracy will stimulate a revolution in the
exploration of stellar and Galactic formation and evolution, and the
determination of the cosmic distance scale (cf.\ \cite{leb00}).

\subsection{Variability}
 
The GAIA large-scale photometric survey will have significant
intrinsic scientific value for stellar astrophysics, providing
valuable samples of variable stars of nearly all types, including
detached eclipsing binaries, contact or semi-contact binaries, and
pulsating stars (cf.\ \cite{pac97}). 
The pulsating stars include key distance calibrators
such as Cepheids and RR~Lyrae stars and long-period variables. Existing
samples are incomplete already at magnitudes as bright as
$V\sim10$~mag. A complete sample of objects will allow determination
of the frequency of variable objects, and will accurately calibrate
period-luminosity relationships across a wide range of stellar
parameters including metallicity.  A systematic variability search will
also allow identification of stars in short-lived but key stages of
stellar evolution, such as the helium core flash and the helium shell
thermal pulses and flashes.  Prompt processing will identify many 
targets for follow-up ground-based studies. Estimated numbers are highly 
uncertain, but suggest 
some 18~million variable stars in total, including 5~million `classic'
periodic variables, 2--3~million eclipsing binaries, 2000--8000 Cepheids,
60\,000--240\,000 $\delta$~Scuti variables, 70\,000 RR~Lyrae, and 
140\,000--170\,000 Miras (\cite{ec00}).

\subsection{Binaries and Multiple Stars}

A key scientific issue regarding double and multiple star formation is the
distribution of mass-ratios $q$. For wide pairs ($>0.5$~arcsec) this
is indirectly given through the distribution of magnitude differences.
GAIA will provide a photometric determination of the $q$-distribution down
to $q\sim0.1$, covering the expected maximum around $q\sim0.2$. 
Furthermore, the large numbers of (`5-year') astrometric orbits, will
allow derivation of the important statistics of the very smallest
(brown dwarf) masses as well as the detailed distribution of orbital
eccentricities (\cite{sod99}).

GAIA is extremely sensitive to non-linear proper motions. A large
fraction of all astrometric binaries with periods from 0.03--30 years
will be immediately recognized by their poor fit to a standard
single-star model. Most will be unresolved, with very unequal
mass-ratios and/or magnitudes, but in many cases a photocentre orbit
can be determined.  For this period range, the absolute and relative
binary frequency can be established, with the important possibility of
exploring variations with age and place of formation in the Galaxy.
Some 10~million binaries closer than 250~pc should be detected, 
with very much larger numbers still detectable out to 1~kpc and beyond.

\subsection{Brown Dwarfs and Planetary Systems}

Sub-stellar companions can be divided in two classes: brown dwarfs 
and planets. There exist three major genesis indicators that can
help classify sub-stellar objects as either brown dwarfs or planets:
mass, shape and alignment of the orbit, and composition and thermal
structure of the atmosphere.  Mass alone is not decisive. The ability
to simultaneously and systematically determine planetary frequency and
distribution of orbital parameters for the stellar mix in the Solar
neighbourhood is a fundamental contribution that GAIA will uniquely
provide. Any changes in  planetary frequency with age or metallicity
will come from observations of stars of all ages

An isolated brown dwarf is typically visible only at ages $<1$~Gyr
because of their rapidly fading luminosity with time.  However, in a binary
system, the mass is conserved, and the gravitational effects on a
main-sequence secondary remain observable over much longer intervals. 
GAIA will have the power to investigate the mass-distribution of
brown-dwarf binaries with 1--30~year periods, of all ages, through
analysis of the astrometric orbits.

There are a number of techniques which in principle allow the
detection of extra-Solar planetary systems: these include pulsar
timing, radial velocity measurements, astrometric techniques, transit
measurements, microlensing, and direct methods based on high-angular
resolution interferometric imaging. A better understanding of the
conditions under which planetary systems form and of their general
properties requires sensitivity to low mass planets (down to
$\sim10M_\oplus$),  characterization of known systems (mass, and
orbital elements), and complete samples of planets, with useful upper
limits on Jupiter-mass planets out to several~AU from the central star
(\cite{mb98b}; \cite{per00}).

Astrometric measurements good to 2--10~$\mu$as will contribute
substantially to these goals, and will complement the ongoing radial
velocity measurement programmes. Although SIM will be able to study in
detail targets detected by other methods, including microlensing,
GAIA's strength will be its discovery potential, following from the
astrometric monitoring of all of the several hundred thousand bright
stars out to distances of $\sim200$~pc (\cite{lss+00}).

\subsection{Solar System}

Solar System objects present a challenge to GAIA because of their
significant proper motions, but they promise a rich scientific reward.
The minor bodies provide a record of the conditions in the proto-Solar
nebula, and their properties therefore shed light on the formation of
planetary systems. 

The relatively small bodies located in the main asteroid belt between
Mars and Jupiter should have experienced limited thermal evolution
since the early epochs of planetary accretion. Due to the radial
extent of the main belt, minor planets provide important information
about the gradient of mineralogical composition of the early
planetesimals as a function of heliocentric distance. It is therefore
important for any study of the origin and evolution of the Solar
system to investigate the main physical properties of asteroids
including masses, densities, sizes, shapes, and taxonomic classes, all
as a function of location in the main belt and in the Trojan clouds.

The possibility of determining asteroid masses relies on the
capability of measuring the tiny gravitational perturbations that
asteroids experience in case of a mutual close approach. At present
only about 10 asteroid masses are known, mostly with quite poor
accuracy.  Asteroid-asteroid encounters have been modelled, and show
that GAIA will allow more than 100 asteroid masses to be determined
accurately.

Albedo is a useful complement to spectrophotometric data for the
definition of different taxonomic classes.  The GAIA photometry will
be much more reliable than most data presently available.  The colour
indices will provide a taxonomic classification for the whole sample
of observed asteroids.

For direct orbit determinations of known asteroids, preliminary
simulations have been performed in which the covariance matrices of
the orbital elements of more than 6000 asteroids were computed using
both the whole set of astrometric observations collected from
ground-based telescopes since 1895 through 1995, as well as a set of
simulated observations carried out by GAIA, computed by considering a
5~year lifetime of the mission, and present instrument performances.
Another set of simulated ground-based observations covering the period
1996--2015 were also performed. For the known asteroids the predicted
ephemeris errors based on the GAIA observations alone 100~years after
the end of the mission are more than a factor~30 better than the
predicted ephemeris errors corresponding to the whole set of past and
future ground-based observations. In other words, after the collection
of the GAIA data, all the results of more than one century of
ground-based asteroid astrometry will be largely superseded.

In addition to known asteroids, GAIA will discover a very large
number, of the order of $10^5$ or $10^6$ new objects, depending on the
uncertainties on the extrapolations of the known population. 
It should be possible to derive precise orbits for many of
the newly discovered objects, since each of them will be 
observed many times during the mission lifetime.  These will include a
large number of near-Earth asteroids. The combination of on-board
detection, faint limiting magnitude, observations at small Sun-aspect
angles, high accuracy in the instantaneous angular velocity
(0.25~mas~s$^{-1}$), and confirmation from successive field transits, means
that GAIA will provide a detailed census of Atens, Apollos and Amors,
extending as close as 0.5~AU to the Sun, and down to diameters of 
about 260--590~m at 1~AU, depending on albedo and observational geometry.

\subsection{Galaxies, Quasars, and the Reference Frame}

GAIA will not only provide a representative census of the stars
throughout the Galaxy, but it will also make unique contributions to
extragalactic astronomy (Table~\ref{tab:gg_local-group}). These
include the structure, dynamics and stellar populations in the Local
Group, especially the Magellanic Clouds, M31 and M33, the space
motions of Local Group galaxies, a multi-colour survey of galaxies
(\cite{vac00}), and studies of supernovae (\cite{hfm99}), galactic
nuclei, and quasars.

\begin{table*} [tbh]
\caption{Local Group galaxies potentially accessible to GAIA.
E(B-V) indicates the foreground reddening, 
and (m-M)$_0$ is the true distance modulus. $V_{\rm lim}$ is the brightest
star in the galaxy. $\mu_{v_t-v_r}$ is the
estimated proper motion, assuming the transverse velocity equals the
observed radial velocity. *~denotes observed values.}
\vspace{0pt}
\begin{center} 
\scriptsize
\begin{tabular}{lrrllccrrr}
\hline 
&&&&&&&&& \\[-5pt]
 Galaxy \qquad\qquad\qquad
	& $l$ & $b$ & $E(B-V)$ & $(m-M)_0$ & Distance 
        & $V_{\rm lim}$ &  N(stars) & $V_r$ & $\mu_{v_t-v_r}$ \\
        & ($^\circ$) & ($^\circ$) & (mag) & (mag)   & (kpc) 
        & (mag)  & (V$<$20)  & (helio) & ($\mu$as/yr) \\[5pt]
\hline 
&&&&&&&&& \\
\noalign{\vskip -5pt}
WLM     &75.9  &$-73.6$ &$0.02\pm0.01$ &$24.83\pm0.08$ &$925\pm40$ 
        &16.5  &$\sim$500 &$-116$ &26 \\
NGC 55  &332.7 &$-75.7$ &$0.03\pm0.02$ &$25.85\pm0.20$ &$1480\pm150$
        &15.0&10's&129&18 \\
IC 10   &119.0 &$-3.3$  &$0.87\pm0.12$ &$24.58\pm0.12$ &$825\pm50$
        &15.0&10's &$-344$&83 \\
NGC 147 &119.8 &$-14.3$ &$0.18\pm0.03$ &$24.30\pm0.12$ &$725\pm45$ 
        &18.5  &10's    &$-193$  &56 \\
And III &119.3 &$-26.2$ &$0.05\pm0.02$ &$24.40\pm0.10$ &$760\pm40$
        &20    &	&              &60 \\
NGC 185 &120.8 &$-14.5$ &$0.19\pm0.02$ &$23.96\pm0.08$ &$620\pm25$
        &20    & 	&$-202$        &69 \\
NGC 205 &120.7 &$-21.7$ &$0.04\pm0.02$ &$24.56\pm0.08$ &$815\pm35$
        &20    &        &$-241$        &62 \\
M32     &121.2 &$-22.0$ &$0.08\pm0.03$ &$24.53\pm0.08$ &$805\pm35$
        &16    &$\sim10^4$ &$-205$     &54 \\
M31     &121.2 &$-21.6$ &$0.08         $&$24.43$       &$770$
        &15    &$\gg10^4$ &$-297$       &81 \\
And I   &121.7 &$-24.9$ &$0.04\pm0.02$ &$24.53\pm0.10$ &$805\pm40$
        &21.7  &        &              & \\
SMC     &302.8 &$-44.3$ &$0.08$        &$18.82$        &$58$   
        &12    &$>10^6$ &158           &900$^*$ \\
Sculptor&287.5 &$-83.2$ &$0.02\pm0.02$ &$19.54\pm0.08$ &$79\pm4$
        &16.0  &100's   &110           &360$^*$ \\
LGS 3   &126.8 &$-40.9$ &$0.08\pm0.03$ &$24.54\pm0.15$ &$810\pm60$
        &      &        &$-277$        &72 \\
IC 1613 &129.8 &$-60.6$ &$0.03\pm0.02$ &$24.22\pm0.10$ &$700\pm35$
        &17.1  &100's   &$-234$        &71 \\
And II  &128.9 &$-29.2$ &$0.08\pm0.02$ &$23.6\pm0.4$   &$525\pm110$
        &20    &        &              & \\
M33     &133.6 &$-31.3$ &$0.08$        &$24.62$        &$840$ 
        &15    &$>10^4$ &$-181$        &46 \\
Phoenix &272.2 &$-68.9$ &$0.02\pm0.01$ &$23.24\pm0.12$ &$445\pm30$
        &17.9  &$\sim10^2$ &56         &27 \\
Fornax  &237.1 &$-65.7$ &$0.03\pm0.01$ &$20.70\pm0.12$ &$138\pm8$
        &14    &100's   &53            &81 \\
EGB0427+63 &144.7 &$-10.5$ &$0.30\pm0.15$ &$25.6\pm0.7$ &$1300\pm700$
        &      &        &$-99$         &16 \\
LMC     &280.5 &$-32.9$ &$0.06$        &$18.45$        &$49$
        &12    &$>10^7$ &278           &1150$^*$ \\
Carina  &260.1 &$-22.2$ &$0.04\pm0.02$ &$20.03\pm0.09$ &$101\pm5$
        &18    &$\sim 10^3$ &229       &478 \\
Leo A   &196.9 &$+52.4$ &$0.01\pm0.01$ &$24.2\pm0.3$   &$690\pm100$
        &      &        &20            &6 \\
Sextans B &233.2 &$+43.8$ &$0.01\pm0.02$ &$25.64\pm0.15$ &$1345\pm100$
        &19.0  &10's    &301           &47 \\
NGC 3109 &262.1 &$+23.1$ &$0.04\pm0.02$ &$25.48\pm0.25$ &$1250\pm165$
        &      &        &403           &68 \\
Antlia  &263.1 &$+22.3$ &$0.05\pm0.03$ &$25.46\pm0.10$ &$1235\pm65$
        &      &        &361           &62 \\
Leo I   &226.0 &$+49.1$ &$0.01\pm0.01$ &$21.99\pm0.20$ &$250\pm30$
        &19    &10's    &168           &142 \\
Sextans A &246.2 &$+39.9$ &$0.03\pm0.02$ &$25.75\pm0.15$ &$1440\pm110$
        &17.5  &10's    &324           &48 \\
Sextans &243.5 &$+42.3$ &$0.03\pm0.01$ &$19.67\pm0.08$ &$86\pm4$
        &      &        &230           &564 \\
Leo II  &220.2 &$+67.2$ &$0.02\pm0.01$ &$21.63\pm0.09$ &$205\pm12$
        &18.6  &100's   &90            &95 \\
GR 8    &310.7 &$+77.0$ &$0.02\pm0.02$ &$25.9\pm0.4$   &$1510\pm330$
        &18.7  &10's    &214           &28 \\
Ursa Minor &105.0 &$+44.8$ &$0.03\pm0.02$ &$19.11\pm0.10$ &$66\pm3$ 
        &16.9  &100's &$-209$          &1000$^*$\\
Draco   &86.4  &$+34.7$ &$0.03\pm0.01$ &$19.58\pm0.15$ &$82\pm6$
        &17    &100's   &$-281$        &1000$^*$ \\
Sagittarius& 5.6 &$-14.1$ &$0.15\pm0.03$ &$16.90\pm0.15$ &$24\pm2$
        &14    &$>10^4$ &140           &2100$^*$ \\
SagDIG  &21.1  &$-16.3$ &$0.22\pm0.06$ &$25.2\pm0.3$   &$1060\pm160$
        &      &        &$-77$         &16 \\
NGC 6822 &25.3 &$-18.4$ &$0.26\pm0.04$ &$23.45\pm0.15$ &$490\pm40$
        &      &        &$-57$         &25 \\
DDO 210 &34.0  &$-31.3$ &$0.06\pm0.02$ &$24.6\pm0.5$   &$800\pm250$
        &18.9  &10's    &$-137         $&36 \\
IC 5152 &343.9 &$-50.2$ &$0.01\pm0.02$ &$26.01\pm0.25$ &$1590\pm200$
        &      &        &124           &16 \\
Tucana  &322.9 &$-47.4$ &$0.00\pm0.02$ &$24.73\pm0.08$ &$880\pm40$
        &18.5  &10's    &              & \\
Pegasus &94.8  &$-43.5$ &$0.02\pm0.01$ &$24.90\pm0.10$ &$955\pm50$
        &20    &        &$-183$        &40 \\[2pt]
\hline 
\end{tabular}
\label{tab:gg_local-group}
\end{center}
\end{table*}

\subsection{The Radio/Optical Reference Frame}

The International Celestial Reference System (ICRS) is realized by the
International Celestial Reference Frame (ICRF) consisting of
212~extragalactic radio-sources with an rms uncertainty in position
between 100 and 500~$\mu$as. The extension of the ICRF to visible
light is represented by the Hipparcos Catalogue. This has rms
uncertainties estimated to be 0.25~mas~yr$^{-1}$ in each component of
the spin vector of the frame, and 0.6~mas in the components of the
orientation vector at the catalogue epoch, J1991.25.  The GAIA
catalogue will permit a definition of the ICRS more accurate by one or
two orders of magnitude than the present realizations (e.g.\ 
\cite{fm98}; \cite{kv99}).

The spin vector can be determined very accurately by means of the many
thousand faint quasars picked up by the astrometric and photometric
survey.  Simulations using realistic quasar counts, conservative
estimates of intrinsic source photocentric instability, and realistic
intervening gravitational lensing effects,  show that an accuracy of
better than $0.4~\mu$as~yr$^{-1}$ will be reached in all three
components of the spin vector. 

For the determination of the frame
orientation, the only possible procedure is to compare the positions
of the radio sources in ICRF (and its extensions) with the positions
of their optical counterparts observed by GAIA. The number of such
objects is currently less than 300 and the error budget is dominated
by the uncertainties of the radio positions.  Assuming current
accuracies for the radio positions, simulations show that the GAIA
frame orientation will be obtained with an uncertainty of
$\sim60~\mu$as in each component of the orientation vector. The actual
result by the time of GAIA may be significantly better, as the number
and quality of radio positions for suitable objects are likely to
increase with time.

The Sun's absolute velocity with respect to a cosmological reference
frame causes the dipole anisotropy of the cosmic microwave background.
The Sun's absolute acceleration can be measured astrometrically: it
will result in the apparent proper motion of quasars. The acceleration of 
the Solar System towards the Galactic centre causes the aberration effect
to change slowly. This leads to a slow change of the apparent
positions of distant celestial objects, i.e., to an apparent proper
motion. For a Solar velocity of 220~km~s$^{-1}$ and a distance of
8.5~kpc to the Galactic centre, the orbital period of the Sun is
$\sim$250~Myr, and the Galactocentric acceleration has the value
0.2~nm~s$^{-2}$, or 6~mm~s$^{-1}$~yr$^{-1}$. A change in velocity by
6~mm~s$^{-1}$ causes a change in aberration of the order of 4~$\mu$as.
The apparent proper motion of a celestial object caused by this effect
always points towards the direction of the Galactic centre. Thus, all
quasars will exhibit a streaming motion towards the Galactic centre of
this amplitude.

\begin{table*}[t]
\caption{Light deflection by masses in the Solar System. The monopole effect
dominates, and is summarized in the left columns for grazing incidence
and for typical values of the angular separation. Columns $\chi_{\rm
min}$ and $\chi_{\rm max}$ give results for the minimum and maximum
angles accessible to GAIA. J$_2$ is the quadrupole moment. The
magnitude of the quadrupole effect is given for grazing incidence, and
for an angle of 1$^\circ$. For GAIA this applies only to Jupiter and
Saturn, as it will be located at L2, with minimum Sun/Earth avoidance
angle of 35$^\circ$.}
\vskip 10pt
\label{tab:light-deflection}
\begin{center}
\leavevmode
\footnotesize
\begin{tabular}{lrrrrrlrr}
\hline 
& \multicolumn{5}{c}{\null} &\multicolumn{3}{c}{\null} \\[-5pt]
Object\qquad\qquad        &\multicolumn{5}{c}{Monopole term} 
                  &\multicolumn{3}{c}{Quadrupole term} \\
 & Grazing &$\chi_{\rm min}$ &$\chi=45^\circ$ &$\chi=90^\circ$
             &$\chi_{\rm max}$ &\qquad \ \ J$_2$ &Grazing &$\chi=1^\circ$ \\
 & ($\mu$as)  & ($\mu$as) & ($\mu$as) & ($\mu$as) & ($\mu$as) 
                                          &   & ($\mu$as)  & ($\mu$as) \\[5pt]
\hline  
&&&&&&&& \\[-5pt]
Sun     &1750000  &13000  &10000 &4100  &2100  &\qquad $\leq 10^{-7}$ & 0.3 & -- \\
Earth   &    500  &    3  &  2.5 & 1.1  &   0  &\qquad 0.001 &  1 & --  \\ 
Jupiter &  16000  &16000  &  2.0 & 0.7  &   0  &\qquad 0.015 &500 &$7\times 10^{-5}$ \\
Saturn  &   6000  & 6000  &  0.3 & 0.1  &   0  &\qquad 0.016 &200 &$3\times 10^{-6}$ \\[5pt]
\hline
\end{tabular}
\end{center}
\end{table*}

\subsection{Fundamental Physics}

The reduction of the Hipparcos data necessitated the inclusion of
stellar aberration up to terms in $(v/c)^2$, and the general
relativistic treatment of light bending due to the gravitational field
of the Sun (and Earth). The GAIA data reduction requires a more
accurate and comprehensive inclusion of relativistic effects, at the
same time providing the opportunity to test a number of parameters of
general relativity in new observational domains, and with much improved
precision.

The dominant relativistic effect in the GAIA measurements is
gravitational light bending, quantified by, and allowing accurate
determination of, the parameter $\gamma$ of the Parametrized
Post-Newtonian (PPN) formulation of gravitational theories. This is of
key importance in fundamental physics. The Pound-Rebka experiment
verified the relativistic prediction of a gravitational redshift for
photons, an effect probing the time-time component of the metric
tensor. Light deflection depends on both the time-space and
space-space components. It has been observed on distance scales of
$10^9-10^{21}$~m, and on mass scales from $1-10^{13} M_\odot$. GAIA
will extend the domain of observations by two orders of magnitude in
length, and six orders of magnitude in mass.

Table~\ref{tab:light-deflection} gives the magnitude of the deflection
for the Sun and the major planets, at different values of the angular
separation $\chi$, for both monopole and quadrupole terms. While
$\chi$ is never smaller than $35^\circ$ for the Sun (a constraint from
GAIA's orbit), grazing incidence is possible for the planets. With the
astrometric accuracy of a few $\mu$as, the magnitude of the expected
effects is considerable for the Sun, and also for observations near
planets. The GAIA astrometric residuals can be tested for any
discrepancies with the prescriptions of general relativity. Detailed
analyses indicate that the GAIA measurements will provide a precision
of about $5\times10^{-7}$ for $\gamma$, based on multiple observations
of $\sim 10^7$ stars with $V<13$~mag at wide angles from the Sun, with
individual measurement accuracies better than $10~\mu$as.

Recent developments in cosmology (e.g.\ inflationary models) and
elementary-particle physics (e.g.\ string theory and Kaluza-Klein
theories), consider scalar-tensor theories as plausible alternatives
to general relativity. A large class of such theories contain an
attractor mechanism towards general relativity in a cosmological
sense; if this is how the Universe is evolving, then today we can
expect discrepancies of the order of $|\gamma-1|\sim10^{-7}-10^{-5}$
depending on the theory.  This kind of argument provides a strong
motivation for any experiments able to reach these accuracies.

\begin{table}[t]
\caption{Perihelion precession due to general relativity and the 
Solar quadrupole moment for a few representative objects. 
$a$~=~semi-major axis; $e$~=~eccentricity; GR~=~perihelion precession 
in mas/yr due to general relativity; J$_2$~=~perihelion precession 
in mas/yr due to the Solar quadrupole moment (assuming J$_2=10^{-6}$).}
\begin{center}
\vspace{5pt}
    \begin{tabular}{lcccc}
\hline &&&& \\[-5pt]
Body & $a$ & $e$& GR &  J$_2$ \\
           &  (AU)& &  (mas/yr) & (mas/yr) \\[5pt]
\hline &&&& \\[-5pt]
 Mercury         & 0.39   & 0.21   &423\phantom{.0} & 1.24\phantom{0}  \\
 Asteroids       & 2.7\phantom{0}  & 0.1\phantom{0}  &\phantom{00}3.4 & 0.001   \\
 1566 Icarus     & 1.08   & 0.83   &102\phantom{.0} & 0.30\phantom{0}  \\
 5786 Talos      & 1.08   & 0.83   &102\phantom{.0} & 0.30\phantom{0}  \\
 3200 Phaeton    & 1.27   & 0.89   &103\phantom{.0} & 0.40\phantom{0}  \\[3pt]
\hline 
\end{tabular}
\label{tab:apollo}
\end{center}
\end{table}

GAIA will observe and discover several hundred thousand minor planets
during its five year mission. Most of these will belong to the
asteroidal main belt, with small orbital eccentricity and semi-major
axes close to 3~AU. The members of the Apollo and Aten groups, which
are all Earth-orbit crossers, will include objects with semi-major
axes of the order of 1~AU and eccentricities as large as~0.9.  The
Amor group have perihelia between 1--1.3~AU, and approach the Earth
but do not cross its orbit.

Relativistic effects and the Solar quadrupole cause the orbital
perihelion of a main belt asteroid to precess at a rate about seven
times smaller than for Mercury in rate per revolution, although more
than a hundred times in absolute rate.

Three cases of Earth-crossing asteroids are considered in
Table~\ref{tab:apollo} giving perihelia precession larger than
Mercury, due to a favorable combination of distance and eccentricity.
The diameters are of the order of 1~km for Icarus and Talos and 4~km
for Phaeton. Observed at a geocentric distance of 1~AU, these objects
have a magnitude between $V=15-17$~mag and an angular diameter of
4~mas and 1~mas respectively. Thus the astrometric measurements will
be of good quality, virtually unaffected by the finite size of the
source. A determination of $\lambda$ with an accuracy of $10^{-4}$ is
a reasonable goal, with a value closer to $10^{-5}$ probably
attainable from the statistics on several tens of planets. An
independent determination of the Solar quadrupole moment J$_2$
requires good sampling in $a(1-e^2)$, and one can expect a result
better than $10^{-7}$.

Revival of interest in the Brans-Dicke-like theories, with a variable
$G$, was partially motivated by the appearance of superstring theories
where $G$ is considered to be a dynamical quantity. Using the white 
dwarf luminosity function an upper bound of $\dot G/G\le -(1\pm
1)\times 10^{-11}$~yr$^{-1}$ has been derived, which is comparable to 
bounds derived  from the binary pulsar PSR~1913+16. Since this is a 
statistical upper limit, any improvement in our knowledge of the white
dwarf luminosity function of the Galactic disk will translate into a   
more stringent upper bound for $\dot G/G$. Since
GAIA will detect numerous white dwarfs at low luminosities, present
errors can be reduced by a factor of roughly~5.  If a reliable age of
the Solar neighbourhood independent of the white dwarf luminosity
function is determinable, the upper limit could be decreased to
$10^{-12}-10^{-13}$~yr$^{-1}$.

\subsection{Summary}

With a census of the accurate positions, distances, space motions
(proper motions and radial velocities), and photometry of all
approximately one billion objects complete to $V=20$~mag, GAIA's 
scientific goals are immense, and can be broadly classified as follows:

The Galaxy: 
origin and history of our Galaxy;
tests of hierarchical structure formation theories; 
star formation history; 
chemical evolution; 
inner bulge/bar dynamics; 
disk/halo interactions; 
dynamical evolution; 
nature of the warp; 
star cluster disruption; 
dynamics of spiral structure; 
distribution of dust;
distribution of invisible mass;
detection of tidally disrupted debris;
Galaxy rotation curve;
disk mass profile. 

Star formation and evolution:
{\it in situ\/} luminosity function;
dynamics of star forming regions; 
luminosity function for pre-main sequence stars; 
detection and categorization of rapid evolutionary phases; 
complete and detailed local census down to single brown dwarfs; 
identification/dating of oldest halo white dwarfs;
age census;
census of binaries and multiple stars.

Distance scale and reference frame: 
parallax calibration of all distance scale indicators; 
absolute luminosities of Cepheids; 
distance to the Magellanic Clouds;
definition of the local, kinematically non-rotating metric. 

Local Group and beyond: 
rotational parallaxes for Local Group galaxies; 
kinematical separation of stellar populations; 
galaxy orbits and cosmological history; 
zero proper motion quasar survey; 
cosmological acceleration of Solar System;
photometry of galaxies; 
detection of supernovae. 

Solar System:
deep and uniform detection of minor planets;
taxonomy and evolution;
inner Trojans;
Kuiper Belt Objects;
near-Earth asteroids;
disruption of Oort Cloud.

Extra-Solar planetary systems:
complete census of large planets to 200--500~pc; 
masses;
orbital characteristics of several thousand systems;
relative orbital inclinations of multiple systems.

Fundamental physics: 
$\gamma$ to $\sim5\times10^{-7}$;
$\beta$ to $3\times10^{-4}-3\times10^{-5}$;
Solar J$_2$ to $10^{-7}-10^{-8}$; 
$\dot G/G$ to $10^{-12}-10^{-13}$~yr$^{-1}$;
constraints on gravitational wave energy for $10^{-12}<f<4\times10^{-9}$~Hz;
constraints on $\Omega_{\rm M}$ and $\Omega_\Lambda$ from quasar 
microlensing.

Examples of specific objects: 
$10^6-10^7$ resolved galaxies; 
$10^5$~extragalactic supernovae;
$500\,000$ quasars;  
$10^5-10^6$ (new) Solar System objects;
$\ga 50\,000$ brown dwarfs;
$30\,000$~extra-Solar planets;
$200\,000$ disk white dwarfs;
$200$ microlensed events;
$10^7$ resolved binaries within 250~pc.

\section{Overall Design Considerations}

Instrument design converges through a consideration of technical
feasibility and scientific requirements. The proposed GAIA design
has arisen from requirements on astrometric precision (10~$\mu$as at 
15~mag), completeness to $V=20$~mag, the acquisition
of radial velocities, the provision of accurate multi-colour
photometry for astrophysical diagnostics, and the need for on-board
object detection (\cite{mig99}; \cite{gbf+00}). 

\subsection{Astrometry}

A space astrometry mission has a unique capability to perform global
measurements, such that positions, and changes in positions caused by
proper motion and parallax, are determined in a reference system
consistently defined over the whole sky, for very large numbers of
objects.  Hipparcos demonstrated that this can be achieved
with milliarcsecond accuracy by means of a continuously scanning
satellite which observes two directions simultaneously. With current
technology this same principle can be applied with a gain of a factor of
more than 100 improvement in accuracy, a factor 1000 improvement in
limiting magnitude, and a factor of $10\,000$ in the numbers of stars
observed.

Measurements conducted by a continuously scanning satellite are
optimally efficient, with each photon acquired during a scan
contributing to the precision of the resulting astrometric parameters.
The over-riding benefit of global astrometry using a scanning
satellite is however not efficiency but reliability: an accurate
instrument calibration is performed naturally, while the
interconnection of observations over the celestial sphere provides the
rigidity and reference system, immediately connected to an
extragalactic reference system, and a realistic determination of the
standard errors of the astrometric parameters. Two individual viewing
directions with a wide separation is the fundamental pre-requisite of
the payload, since this leads to the determination of absolute
trigonometric parallaxes, and absolute distances, exploiting the
method implemented for the first time in the Hipparcos mission.

The ultimate accuracy with which the direction to a point source of
light can be determined is set by the dual nature of electromagnetic
radiation, namely as waves (causing diffraction) and particles
(causing a finite signal-to-noise ratio in the detection process). For
wavelength $\lambda$ and telescope aperture $D$ the characteristic
angular size of the diffraction pattern image is of order $\lambda/D$
radians.  If a total of $N$ detected photons are available for
localizing the image, then the theoretically achievable angular
accuracy will be of order $(\lambda/D)\times N^{-1/2}$ radians. A
realistic size for non-deployable space instruments is of order
2~m. Operating in visible light ($\lambda \sim 0.5~\mu$m) then
gives diffraction features of order $\lambda/D \sim 0.05$~arcsec.  To
achieve a final astrometric accuracy of 10~$\mu$as it is therefore
necessary that the diffraction features are localised to within
$1/5000$ of their characteristic size. Thus, some 25~million detected
photons are needed to overcome the statistical noise, although
extreme care will be needed to achieve such precision in practice. The
requirement on the number of photons can be satisfied for objects
around 15~mag with reasonable assumptions on collecting area and
bandwidth.  Quantifying the tradeoff between dilute versus filled
apertures, allowing for attainable focal lengths, attainable pixel
sizes, component alignment and stability, and data rates, has clearly
pointed in the direction of a moderately large filled aperture (as
apposed to an interferometric design) as the optical system of choice.

The GAIA performance target is 10~$\mu$as at 15~mag. Restricting GAIA
to a limiting magnitude of 15~mag, or to a subset of all objects down to
its detection limit, would provide a reduction in the down-link
telemetry rate, but little or no change in the other main aspects of
the payload design. These are driven simply by the photon noise budget
required to reach a 10~$\mu$as accuracy at 15~mag. The faint magnitude
limit, the ability to meet the adopted scientific case, and the number
of target objects follow from the accuracy requirement, with no
additional spacecraft cost.

\subsection{Radial Velocity Measurements}

There is one dominant scientific requirement, as well as two additional
scientific motivations, for the acquisition of radial velocities with
GAIA: (i) astrometric measurements supply only two components of the
space motion of the target stars: the third component, radial
velocity, is directed along the line of sight, but is nevertheless
essential for dynamical studies; (ii) measurement of the radial
velocity at a number of epochs is a powerful method for detecting and
characterising binary systems; (iii) at the GAIA accuracy levels, 
`perspective acceleration' is at the same time both a complication and
an important observable quantity. If the distance between an object
and observer changes with time due to a radial component of motion, a
constant transverse velocity is observed as a varying transverse
angular motion, the perspective acceleration. Although the effect is
generally small, some hundreds of thousands of high-velocity stars will have
systematic distance errors if the radial velocities are unknown.

On-board acquisition of radial velocities with GAIA is not only
feasible, but is relatively simple, is scientifically necessary, and
cannot be readily provided in any other way. In terms of accuracy
requirements, faint and bright magnitude regimes can be distinguished.
The faint targets will mostly be distant stars, which will be of
interest as tracers of Galactic dynamics. The uncertainty in the
tangential component of their space motion will be dominated by the
error in the parallax. Hence a radial velocity accuracy of $\simeq
5$~km~s$^{-1}$ is sufficient for statistical purposes.  Stars with
$V\la15$~mag will be of individual interest, and the radial velocity
will be useful also as an indicator of multiplicity and for the
determination of perspective acceleration. The radial velocities will 
be determined by digital cross-correlation between an observed spectrum
and an appropriate template. The present design allows (for red
Population~I stars of any luminosity class) determination of radial
velocities to $\sigma_v \simeq 5$~km~s$^{-1}$ at $V=18$~mag 
(e.g.\ \cite{mun99b}).

Most stars are intrinsically red, and made even redder by interstellar
absorption.  Thus, a red spectral region is to be preferred for the
GAIA spectrograph. To maximize the radial velocity signal even for
metal-poor stars, strong, saturated lines are desirable. Specific
studies, and ground-based experience, show that the Ca\,{\sc ii} triplet
near 860~nm is optimal for radial velocity determination in the
greatest number of stellar types.

Ground-based radial velocity surveys are approaching the one
million-object level. That experience shows the cost and complexity of
determining some hundreds of millions of radial velocities is
impractical.  There is also a substantial additional scientific return
in acquiring a large number of measurements, and doing so not only
well spaced in time but also, preferably, simultaneously with the
astrometric measurements (e.g.\ variables and multiple systems).

\subsection{Derivation of Astrophysical Parameters}

The GAIA core science case requires measurement of luminosity,
effective temperature, mass, age and composition, in addition to
distance and velocity,  to optimise understanding of the stellar
populations in the Galaxy and its nearest neighbours.  The quantities
complementary to the kinematics can be derived from the spectral
energy distribution of the stars by  multi-band photometry and
spectroscopy. Acquisition of this astrophysical information is an
essential part of the GAIA payload. A broad-band magnitude, and its
time dependence, will be obtained from the primary mission data,
allowing both astrophysical analyses and the critical corrections for
residual system chromaticity. For the brighter stars, the radial
velocity spectra will complement the photometric data.

\begin{figure*}[t]
  \begin{center}
    \leavevmode
\centerline{
\epsfig{file=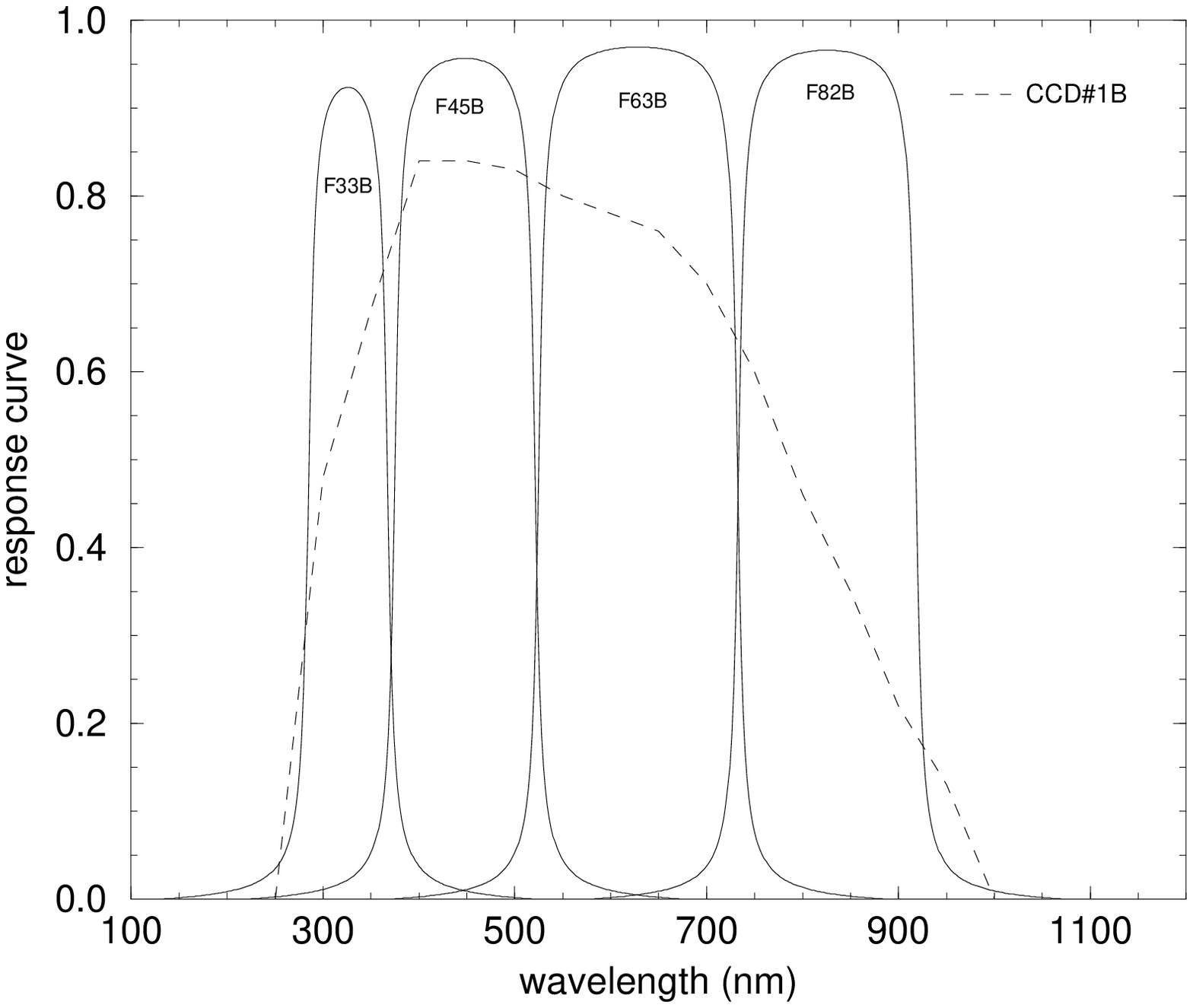,height=5.6cm,angle=0}
\epsfig{file=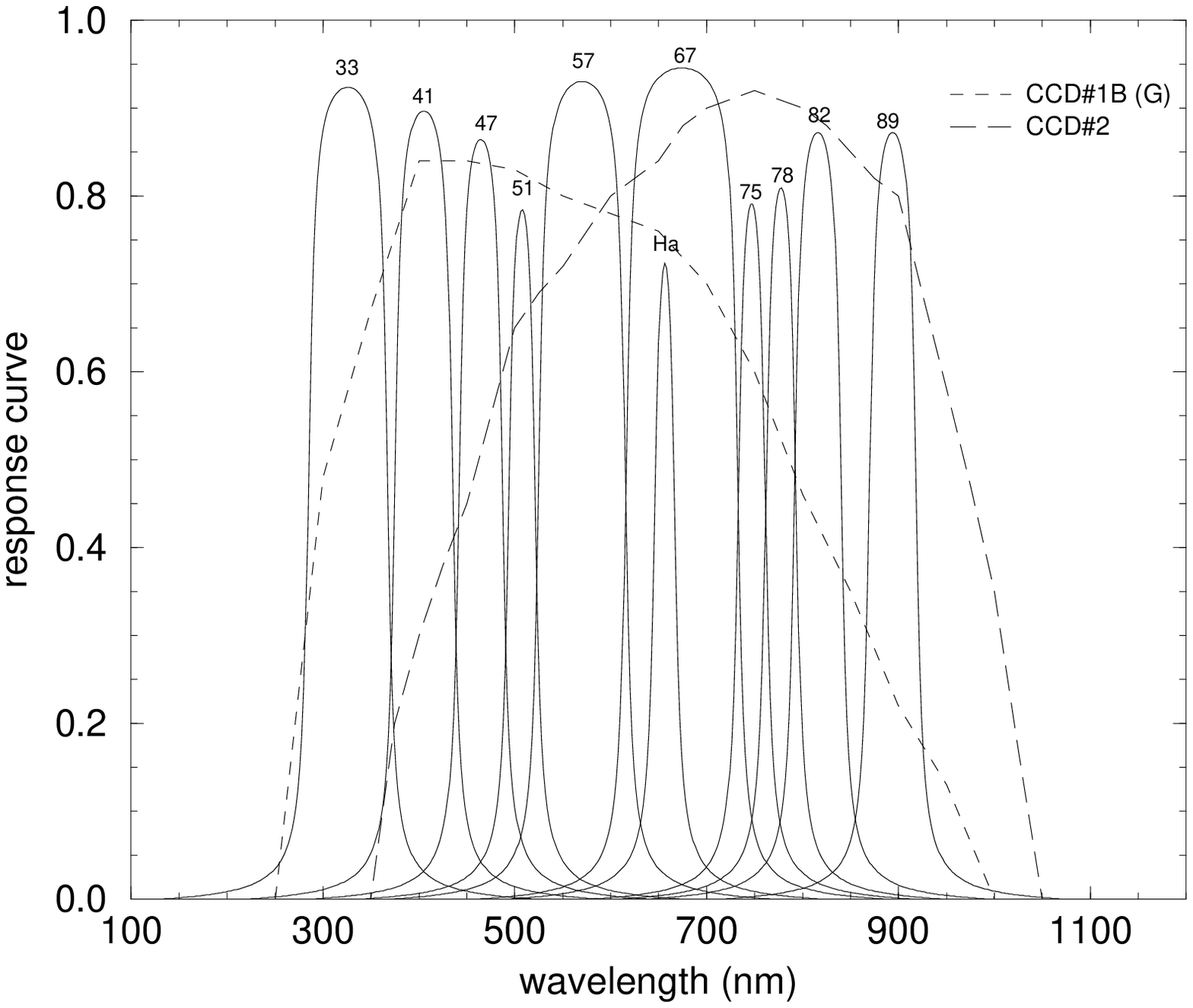,height=5.6cm,angle=0}
}
  \end{center}
  \vskip -15pt
\caption{Filter transmission curves and CCD response curves 
for the provisional (baseline) broad-band (left) and medium-band
(right) photometric systems.}
\label{fig:photometric-system}
\end{figure*}

For essentially every application of the GAIA astrometric data,
high-quality photometric data will be crucial, in providing the basic
tools for classifying stars across the entire HR~diagram, as well as
in identifying specific and peculiar objects (e.g.\ \cite{str99}).  Photometry 
must determine (i) temperature and reddening at least for OBA stars and
(ii) effective temperatures and abundances for late-type giants and
dwarfs.  To be able to reconstruct Galactic formation history the
distribution function of stellar abundances must be determined to
$\sim 0.2$~dex, while effective temperatures must be determined to
$\sim 200$~K. Separate determination of the abundance of Fe and
$\alpha$-elements (at the same accuracy level) will be desirable for
mapping Galactic chemical evolution.  These requirements translate
into a magnitude accuracy of $\simeq 0.02$~mag for each colour index.

Many photometric systems exist, but none is necessarily optimal for space
implementation.  For GAIA, photometry will be required for quasar and
galaxy photometry, Solar System object classification, etc.
Considerable effort has therefore been devoted to the design of an
optimum filter system for GAIA (e.g.\ \cite{hfk+99}; \cite{mun99a}).  
The result of this effort is a
baseline system, with four broad and eleven medium passbands, covering
the near ultraviolet to the CCD red limit. The filters are summarised in
Figure~\ref{fig:photometric-system}. The 4~broad-band filters are 
implemented within the astrometric fields, and therefore yield photometry 
at the same angular resolution (also relevant for chromatic 
correction), while the 11~medium-band filters are implemented within 
the spectrometric telescope. Both target magnitude limits of 20~mag, 
as for the astrometric measurements.

\subsection{On-Board Detection}

Clear definition and understanding of the selection function used to
decide which targets to observe is a crucial scientific issue,
strongly driving the final scientific output of the mission. The
optimum selection function, and that adopted, is to detect every
target above some practical signal level on-board as it enters the
focal plane.  This has the advantage that the detection will be
carried out in the same wave-band, and at the same angular resolution,
as the final observations.  The focal plane data on all objects down
to about 20~mag can then be read out and telemetered to ground within
system capabilities. All objects, including Solar System objects, 
variable objects, supernovae, and microlensed sources, are detected using 
this `astrometric sky mapper', described in further detail in 
Section~\ref{sec:focal-plane}.

\begin{figure}[t]
\begin{center}
\leavevmode
\centerline{\epsfig{file=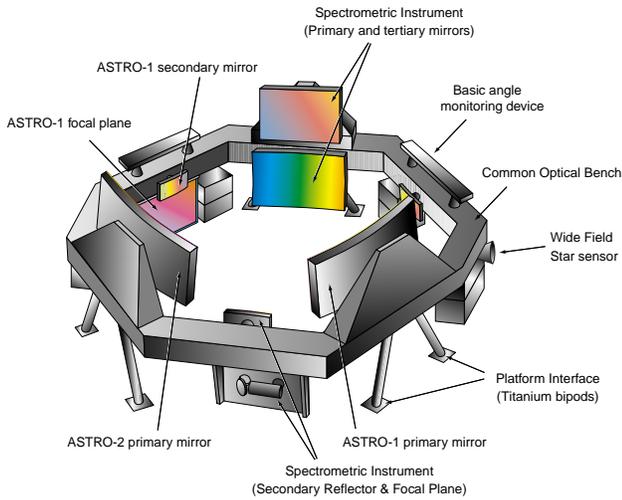,width=8.8cm,angle=0,
	bbllx=20,bblly=160,bburx=560,bbury=640,clip=}}
\end{center}
\vskip -20pt
\caption{The payload includes two identical astrometric instruments 
(labelled ASTRO-1 and ASTRO-2) separated by the 106$^\circ$ basic angle, 
as well as a spectrometric instrument (comprising a radial velocity 
measurement instrument and a medium-band photometer) which share the 
focal plane of a third viewing direction. All telescopes are 
accommodated on a common optical bench of the same material, and a 
basic angle monitoring device tracks any variations in the relative 
viewing directions of the astrometric fields. }
\label{fig:payload}
\end{figure}

\section{Payload Design}

\subsection{Measurement Principles}

The overall design constraints have been investigated in detail in 
order to optimise the number and optical design of each viewing direction, the
choice of wavelength bands, detection systems, detector sampling
strategies, basic angle, metrology system, satellite layout, and
orbit (\cite{msc+99}). The resulting proposed payload design
(Figure~\ref{fig:payload}) consists of:

(a) two astrometric viewing directions. Each of these astrometric
instruments comprises an all-reflective three-mirror telescope with an
aperture of $1.7\times0.7$~m$^2$, the two fields separated by a basic
angle of 106$^\circ$. Each astrometric field comprises an astrometric
sky mapper, the astrometric field proper, and a broad-band
photometer. Each sky mapper system provides an on-board
capability for star detection and selection, and for the star position
and satellite scan-speed measurement. The main focal plane assembly
employs CCD technology, with about 250~CCDs and accompanying video
chains per focal plane, a pixel size 9~$\mu$m along scan, TDI
(time-delayed integration) operation, and an integration time of 
$\sim0.9$~s per CCD;

(b) an integrated radial velocity spectrometer and photometric
instrument, comprising an all-reflective three-mirror
telescope of aperture $0.75\times0.70$~m$^2$. The field of view is
separated into a dedicated sky mapper, the radial velocity
spectrometer, and a medium-band photometer. Both
instrument focal planes are based on CCD technology operating in TDI
mode;

(c) the opto-mechanical-thermal assembly comprising: (i) a single
structural torus supporting all mirrors and focal planes, employing
SiC for both mirrors and structure. There is a symmetrical
configuration for the two astrometric viewing directions, with the
spectrometric telescope accommodated within the same structure, between the two
astrometric viewing directions; (ii) a deployable Sun shield to avoid
direct Sun illumination and rotating shadows on the payload module,
combined with the Solar array assembly; (iii) control of the heat
injection from the service module into the payload module, and control
of the focal plane assembly power dissipation in order to provide an
ultra-stable internal thermal environment; (iv) an alignment mechanism
on the secondary mirror for each astrometric instrument, with
micron-level positional accuracy and 200~$\mu$m range, to correct for
telescope aberration and mirror misalignment at the beginning of life;
(v) a permanent monitoring of the basic angle, but without active
control on board.

The accuracy goal is to reach a 10~$\mu$as rms positional accuracy for
stars of magnitude $V=15$~mag. For fainter magnitudes, the accuracy
falls to about $20-40~\mu$as at $V=17-18$~mag, and to $100-200~\mu$as
at $V=20$~mag, entirely due to photon statistics. For $V<15$~mag,
higher accuracy is achieved, but will be limited by systematic effects
at about $3-4~\mu$as for $V<10-11$~mag. Raw data representing the star
profile along scan must be sent to ground. An integral objective of the
mission is to provide the sixth astrometric parameter, radial
velocity, by measuring the Doppler shift of selected spectral lines.
Colour information is to be acquired for all observed objects,
primarily to allow astrophysical analyses, though calibration of the
instrument's chromatic dependence is a key secondary consideration.

The astrometric accuracy can be separated into two independent terms,
the random part induced by photo-electron statistics on the
localisation process accuracy, and a bias error which is independent
of the number of collected photons. The random part decreases in an
ideal system as $N^{-0.5}$, where $N$ is the number of detected
electrons per star; the bias part is independent of $N$, represents
the ultimate capability of the system for bright stars, is limited by
payload stability on timescales shorter than those which can be
self-calibrated, i.e.\ shorter than about 5~hours.

GAIA will operate through continuous sky scanning, this mode being
optimally suited for a global, survey-type mission with very many
targets, and being of proven validity from Hipparcos. The satellite
scans the sky according to a pre-defined pattern in which the axis of
rotation (perpendicular to the three viewing directions) is kept at a
nominally fixed angle $\xi$ from the Sun, describing a precessional
motion about the Solar direction at constant speed with respect to the
stars. This angle is optimised against satellite Sun shield demands,
parallax accuracy, and scanning law. Resulting satellite pointing
performances are determined from operational and scientific processing
requirements on ground, and are summarised in
Table~\ref{tab:pointing}. 

A mission length of 5~years is adopted for the satellite design
lifetime, which starts at launcher separation and includes the
transfer phase and all provisions related to system, satellite or
ground segment dead time or outage. A lifetime of 6~years has been
used for the sizing of all consumables.

\begin{table}[t]
\caption{Summary of the scanning law and pointing requirements.
0.05~Hz is the maximum frequency that can be identified after 
measurement post-processing.}
\label{tab:pointing}
\begin{center}
\footnotesize
\begin{tabular}{ll}
\hline & \\[-5pt]
Parameter&      Value \\[5pt]
\hline &\\[-5pt]
Satellite scan axis tilt angle& 55$^\circ$ to the Sun  \\
Scan rate& 120 arcsec~s$^{-1}$ \\
Absolute scan rate error& 1.2 arcsec~s$^{-1}$ (3$\sigma$) \\
Precession rate&   0.17 arcsec~s$^{-1}$ \\
Absolute precession rate error& 0.1 arcsec~s$^{-1}$ (3$\sigma$) \\
Absolute pointing error&  5 arcmin (3$\sigma$) \\
Attitude absolute measurement error&  0.001 arcsec (1$\sigma$) \\
High-frequency disturbances: & \\
\multicolumn{1}{r}{power spectral density at 0.05~Hz}&
$\le1000~\mu$as$^2$ Hz$^{-1}$ \\ 
\multicolumn{1}{r}{for $f > 0.05$~Hz}& decreasing as $f^{-2}$ \\
[5pt]
\hline 
\end{tabular}  
\end{center}
\end{table}

\subsection{Optical Design}

The astrometric telescopes have a long focal length, necessary for
oversampling the individual images. A pixel size of
9~$\mu$m in the along-scan direction was selected, with the 50~m focal
length allowing a 6-pixel sampling of the diffraction image along scan
at 600~nm. The resulting optical system is very compact, fitting into
a volume 1.8~m high, and within a mechanical structure adapted to the
Ariane~5 launcher. Deployable payload elements have been avoided.
System optimisation yields a suitable full pupil of
$1.7\times0.7$~m$^2$ area with a rectangular shape. Optical
performances which have been optimised are the image quality,
characterised by the wave-front error, and the along-scan
distortion, avoiding at the same time a curved focal plane in order to
facilitate CCD positioning and mechanical complexity. The optical
configuration is derived from a three-mirror anastigmatic design with
an intermediate image. The three mirrors have aspheric surfaces with
limited high-order terms, and each of them is a part of a rotationally
symmetric surface. The aperture shapes are rectangular and decentered,
while each mirror is slightly tilted and decentered.

The tolerable optical distortion arises from the requirement that any
variation of scale across the field must not cause significant image
blurring during TDI operation. The number and size of the CCDs has
been determined to match the optical quality locally in the field.

The monochromatic point spread function, $P_\lambda(\xi,\eta)$, at a
specific point in the field, is related to the corresponding wavefront
error map $w(x,y)$ in the pupil plane through the diffraction formula.
The overall wavefront error of the telescope is the sum of the errors
arising from optical design, alignment, and polishing residuals 
for the three mirrors. The design target of $\lambda/50$~rms over the
whole field corresponds to a Strehl ratio of 0.84 at 500~nm.
From analysis performed using the optical design software package
Code~V, alignment errors can be made negligible (wave front error
$<\lambda$/70 rms) provided that the mirrors are positioned with an
accuracy of about $\pm1~\mu$m. A 5~degree-of-freedom compensation
mechanism with this accuracy (not considered to be excessively
stringent with piezo-type actuators) is therefore implemented on the
secondary reflector of each of the astrometric telescopes. This allows 
optimization of the overall optical quality
in orbit as a result of on-ground residual alignment errors, and the 
recovery of misalignments of the telescope optics which may be induced by
launch effects, even if all the mirrors are randomly misaligned by an
amplitude $\pm50~\mu$m in all directions. The required wavefront error
measurement will be performed on at least three points of the field of
view. In practice, the astrometric performance is not strongly
dependent on the actual telescope wavefront error, since the effect of
aberrations corresponds to first order in an energy loss in the
central diffraction peak, which is the only part of the point spread
function used for the star localization.

Although the optical design only employs mirrors, diffraction effects
with residual (achromatic) aberrations induce a small chromatic shift
of the diffraction peak. The chromaticity image displacement depends
on position in the field, and on the star's spectral energy
distribution (colour), but not on its magnitude. One purpose of the
broad-band photometric measurements within the main field is to provide 
colour information on each
observed object in the astrometric field to enable this chromaticity
bias calibration on ground. Recent developments made on ion beam
polishing have shown that polishing errors can be made practically 
negligible ($\lambda$/100 rms obtained on a SiC reflector of about
200~mm diameter). It is therefore likely that the chromatic shift can
be reduced below a few tens of $\mu$as over the whole field, easing
calibration requirements.
Combination of the satellite Sun shield and internal baffling reduce
straylight to negligible levels.

\begin{figure}[t]
{\epsfig{file=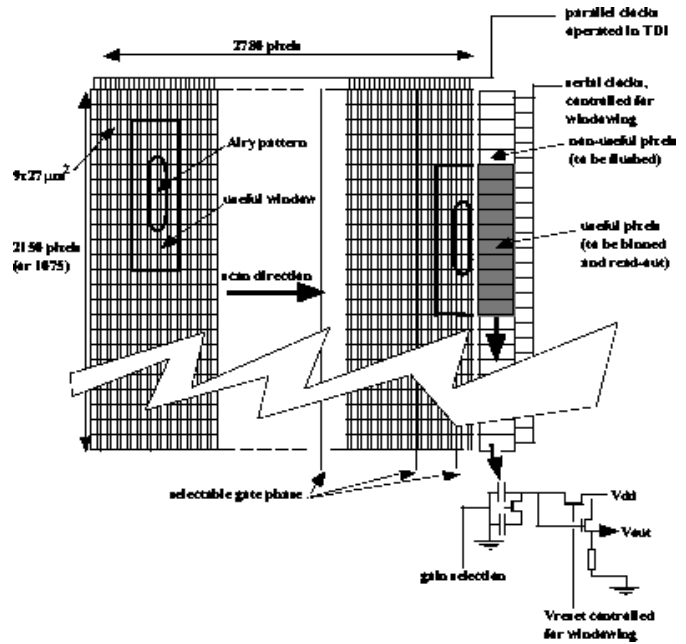,width=8.8cm}}
\caption{Operating mode for the astrometric field CCDs. The 
location of the star is known from the astrometric sky mapper, 
combined with the satellite attitude, a window is selected around the 
star in order to minimise the resulting read-out noise of the relevant 
pixels.}
\end{figure}

\subsection{Astrometric Focal Plane}
\label{sec:focal-plane}

The focal plane contains a set of CCDs operating in TDI
(time-delayed integration) mode, scanning at the same velocity as the
spacecraft scanning velocity and thus integrating the stellar images
until they are transferred to the serial register for read out. Three
functions are assigned to the focal plane system: (i) the astrometric
sky mapper; (ii) the astrometric field, devoted to the
astrometric measurements; (iii) the broad band photometer, which
provides broad-band photometric measurements for each object. The same
elementary CCD is used for the entire focal plane, with minor
differences in the operating modes depending on the assigned functions.

The astrometric sky mapper detects objects entering the field of view,
and communicates details of the star transit to the subsequent
astrometric and broad-band photometric fields. Three CCD strips
provide (sequentially) a detection region for bright stars, a region 
which is read out completely to detect all objects crossing the field, 
and a third region which reads out detected objects in a windowed mode, 
to reduce read-out noise (and hence to improve the signal-to-noise 
ratio of the detection process), and to confirm objects provisionally 
detected in the previous CCD strips, in the presence of, e.g., cosmic rays.
Simulations have shown that algorithms such as those developed for the 
analysis of crowded photometric fields (e.g.\ \cite{irw85}) can be 
adapted to the problem of on-board detection, yielding good detection 
probabilities to 20~mag, with low spurious detection rates. Passages 
of stars across the sky mapper yield the instantaneous satellite spin 
rate, and allow the prediction of the the individual star transits 
across the main astrometric field with adequate precision for the 
foreseen windowing mode.

The size of the astrometric field is optimised at system level to
achieve the specified accuracy, with a field of
0\ddeg5~$\times$~0\ddeg66.  The size of the individual CCD is a
compromise between manufacturing yield, distortion, and integration
time constraints. The pixel size is a compromise between manufacturing
feasibility, detection performances (QE and MTF), and charge-handling
capacity: a dimension of 9~$\mu$m in the along-scan direction provides
full sampling of the diffraction image, and a size of 27~$\mu$m in the
across-scan direction is compatible with the size of the dimensions of
the point spread function and cross scan image motion. In addition, it
provides space for implementation of special features for the CCD
(e.g.\ pixel anti-blooming drain) and provides improved
charge-handling capacity.  Quantitative calculations have demonstrated
that the pixel size, TDI smearing, pixel sampling, and point spread
function are all matched to system requirements. The CCDs are slightly
rotated in the focal plane and are individually sequenced in order to
compensate for the telescope optical distortion. Cross-scan binning of
8~pixels is implemented in the serial register for improvement of the
signal-to-noise ratio. 

\begin{figure}[t]
\begin{center}
\leavevmode
\centerline{\epsfig{file=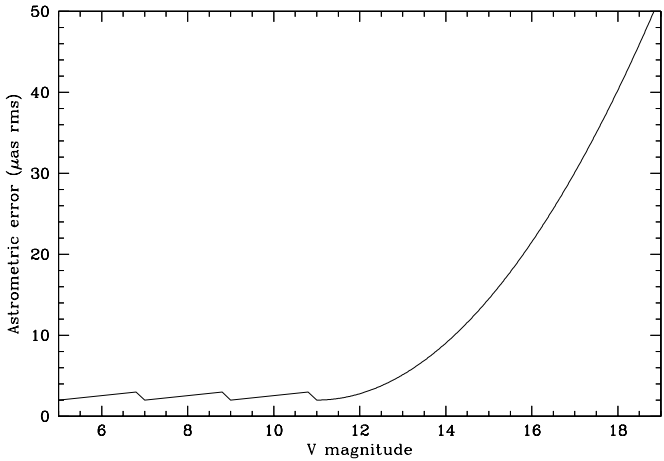,width=8.8cm}}
\end{center}
\vskip -10pt
\caption{Nominal accuracy performance versus magnitude, for a G2V~star. 
For $V<$~15~mag the astrometric performance 
improves because the number of detected photons increases
until the detector saturation level is reached 
(\/$V\sim$~11.6 for G2V star). Brighter than this, 
the performance is practically independent of magnitude, due to the 
pre-selection of the number of TDI stages required to avoid saturation.}
\label{fig:ccd-accuracy}
\end{figure}

Each individual CCD features specific architecture allowing
measurement of stars brighter than the normal saturation limit of
about $V=11-12$~mag: selectable gate phases allow pre-selection of the
number of TDI stages to be used within a given CCD array. The
resulting astrometric error versus magnitude shows the effect of this
discrete selection (Figure~\ref{fig:ccd-accuracy}).

At the apparent magnitude and integration time limits appropriate for
GAIA most of the pixel data do not include any useful information.
There is a clear trade-off between reading too many pixels, with
associated higher read-noise and telemetry costs, and reading too few,
with associated lost science costs.  This contributes to the choice of
on-board real-time detection, with definition of a window around each
source which has sufficient signal to be studiable, and determination
of the effective sensitivity limit to be that which saturates the
telemetry, and which provides a viable lower signal. Combining all
these constraints sets the limit near $V=20$~mag, resulting in an 
estimated number of somewhat over one billion targets.

The broad-band photometric field provides multi-colour, multi-epoch
photometric measurements for each object observed in the astrometric
field, for chromatic correction and astrophysical analysis.  Four
photometric bands are implemented within each instrument.

\subsection{Spectrometric Instrument}

A dedicated telescope, with a rectangular entrance pupil of
$0.75\times0.70$~m$^2$, feeds both the radial velocity spectrometer
and the medium-band photometer: the overall field of view is split
into a central $1^\circ\times1^\circ$ devoted to the radial velocity
measurements, and two outer $1^\circ\times1^\circ$ regions devoted to
medium-band photometry. The telescope is a 3-mirror standard
anastigmatic of focal length  4.17~m. The mirror surfaces are coaxial
conics. An all-reflective design allows a wide spectral bandwidth for
photometry. The image quality at telescope focus allows the use of
$10\times10~\mu$m$^2$ pixels within the photometric field,
corresponding to a spatial resolution of 0.5~arcsec. 

\begin{figure*}[t]
\begin{center}
\leavevmode
\centerline{\epsfig{file=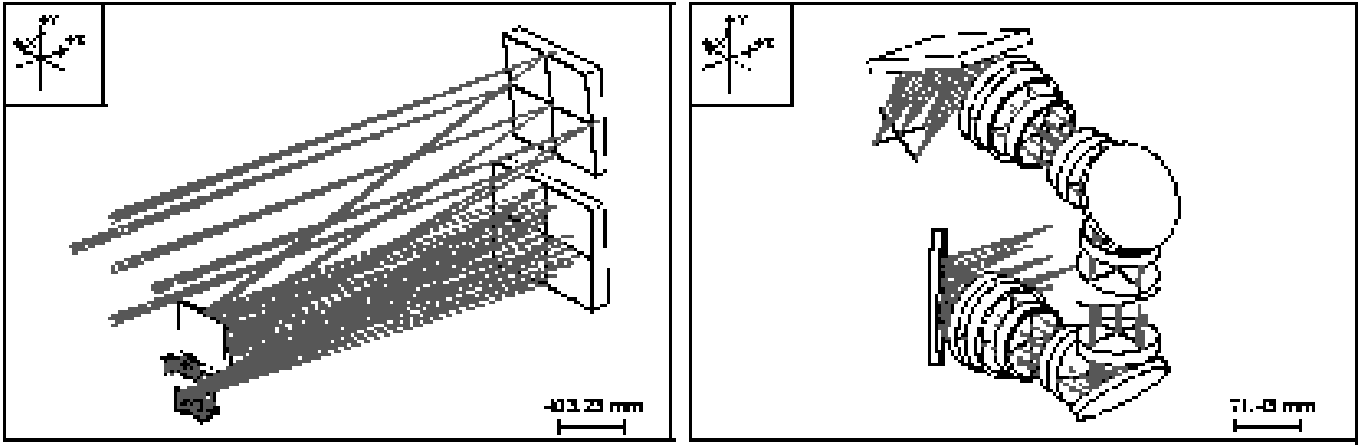,width=18.0cm}}
\end{center}
\vskip -10pt
\caption{Optical configuration of the spectrometric instrument. The 
left figure shows the overall telescope design, while the right figure 
shows details of the spectrograph optics.}
\label{fig:spectro-optics}
\end{figure*}

The radial velocity spectrometer acquires spectra of all sufficiently
bright sources, and is based on a slit-less spectrograph comprising a
collimator, transmission grating plus prism (allowing TDI operation
over the entire field of view) and an imager, working at unit
magnification. The two lens assemblies (collimating and focusing) are
identical, compensating odd aberrations including coma and distortion.
The dispersion direction is perpendicular to scan direction. The
overall optical layout is shown in Figure~\ref{fig:spectro-optics}.
The array covers a field height of 1$^\circ$. Each
$20\times20~\mu$m$^2$ pixel corresponds to an angular sampling of
1~arcsec and a spectral sampling of approximately 0.075~nm/pixel. The
focal plane consists of three CCDs mechanically butted together, each
operated in TDI mode with its own sequencing, providing read-noise as
low as 3~e$^-$ rms with the use of a dedicated `skipper-type' multiple
non-destructive readout architecture with 4~non-destructive readout
samples per pixel.

The requirements for the CCDs are very similar to those of the
astrometric field, including the use of TDI, and dedicated sky mapper.
The photometric bands (Figure~\ref{fig:photometric-system})
will require filters to be directly fixed onto the CCD array.

\subsection{Science Data Acquisition and On-Board Handling}

Preliminary investigations have been carried out to identify the minimum 
set of data to be transmitted to ground to satisfy the 
scientific mission objectives; to identify some on-board data
discrimination compression principles able to provide the targeted
data compression ratio; to assess the feasibility and complexity of
implementing such compression strategies and related algorithms on
board; to assess the resulting compressed data rate at payload output,
which are used for the sizing of the solid state memory and 
communication subsystem; and to derive preliminary mass, size, and
power budgets for the on-board processing hardware. For estimating 
telemetry rates (Table~\ref{tab:telemetry}), a specific spatial sampling 
of the CCD data has been assumed. This sampling is not yet
optimised and final, but represents a useful first approximation.

The instantaneous data rate will primarily
fluctuate with the stellar density in each of the three fields of
view, which scale with Galactic latitude.  On-board storage will store
a full day of observation for downlink at a higher rate during
ground-station visibility. 
Including overhead, the total raw science data rate is roughly a
factor 7 higher than the mean (continuous) payload data rate foreseen
in the telemetry budget ($\sim 1$~Mbit~s$^{-1}$).  Data compression
will reduce this discrepancy, but there remains roughly a
factor two to be gained either by smarter CCD
sampling, or by increasing the link capacity.

\subsection{CCD Details}

CCD detectors form the core of the GAIA payload: their
development and manufacture represents one of the key challenges 
for the programme. 
In the present study, consideration was given to the 
requirements on electro-optical behaviour; array size; buttability; 
pixel size; bright star handling; serial register performance; 
output amplifiers; power dissipation in the image zone, serial 
register, and output amplifier; trade-off between QE and MTF;
photo-response non-uniformity; dark current; conversion factor and 
linearity; charge handling capacity per pixel; charge transfer 
efficiency in the image zone and serial register; minimization of 
residual images; anti-blooming efficiency; and packaging. The present 
baseline design is summarised in Table~\ref{tab:ccd}.

For astrometric use CCD accuracy
depends essentially on the integral of QE$\times$MTF over the
wavelength band. CCD MTF must be optimised in parallel with the
QE. The QE and MTF values for the CCD optimised for the astrometric
field have been used in the detailed astrometric accuracy analysis.
The pixel size, nominally adopted as $9\times27~\mu$m$^2$ for the 
astrometric field, is an important design parameter.  A smaller
pixel size would decrease the telescope focal length, as well as the 
overall size of the astrometric focal plane assembly, and consequently
the overall size of the overall payload. However, such devices provide 
worse performance in a number of other areas. The trade-off 
between QE, MTF, and charge-handling capacity results in 
design reference values, and bread-boarding
activities are underway to verify these performances in detail.

\begin{table*}[t]
\begin{center}
\caption{Average stellar flow in the various fields of the
astrometric and spectrometric instruments, and the resulting 
average telemetry rates.  A limiting of $G=20$~mag is
assumed for the astrometric instrument (AF and BBP) and for the
medium-band photometer (MBP), and $G=17$~mag for the radial-velocity
spectrometer (RVS).  It is assumed that each sample 
represents 16~bits of raw data.  The resulting raw data rates
are before compression and do not include overhead.}
\vskip 10pt
    \leavevmode    
    \footnotesize
    \begin{tabular}[tbh]{l|cccc|ccc}
\hline &&&&&&& \\[-4pt]
Parameter 
  & \multicolumn{4}{c|}{Astro-1 and 2 (per instrument)} 
  & \multicolumn{3}{c}{Spectro} \\
& ASM3 & AF01--16 & AF17 & BBP & SSM1 & MBP & RVS \\[4pt]
\hline &&&&&&& \\[-4pt]
Limiting magnitude, $G_{\rm max}$ [mag] 
  & \multicolumn{4}{c|}{20} 
  & \multicolumn{2}{c}{20} 
  & \multicolumn{1}{c}{17} \\
Average star density, $N_{\rm s}$ [deg$^{-2}$] 
  & \multicolumn{4}{c|}{25\,000} 
  & \multicolumn{2}{c}{25\,000} 
  & \multicolumn{1}{c}{2900} \\
TDI integration time per CCD, $\tau_1$ [s] 
  & \multicolumn{4}{c|}{0.86} 
  & \multicolumn{2}{c}{3.0} 
  & \multicolumn{1}{c}{30} \\
Field width across scan, $\Phi_y$ [deg] 
  & \multicolumn{4}{c|}{0.66} 
  & \multicolumn{2}{c}{1.0} 
  & \multicolumn{1}{c}{1.0} \\
Star flow through $\Phi_y$, $f=N_{\rm s}\Phi_y\omega$ [s$^{-1}$] 
  & \multicolumn{4}{c|}{550} 
  & \multicolumn{2}{c}{833} 
  & \multicolumn{1}{c}{97} \\
Number of CCDs along scan, $N_{\rm CCD}$ 
  & 1 & 16 & 1 & 4 & 1 & 14 & 1 \\
Solid angle of CCDs, $\Omega$ [deg$^2$] 
  & 0.019 & 0.302 & 0.019 & 0.077 & 0.100 & 1.400 & 1.000 \\
Number of stars on the CCDs, $N_{\rm s}\Omega$ 
  & 473 & 7568 & 473 & 1892 & 2500 & 35\,000 & 2900 \\
Readout rate, $R=N_{\rm s}\Omega/\tau_1$ [s$^{-1}$]
  & 550 & 8800 & 550 & 2200 & 833 & 11\,662 & 97 \\
Samples per star read out
  & 25 & 6 & 30 & 16 or 10 & 42 & 14 & 930 \\
Samples per star transmitted, $n$
  & 25 & 6 & 30 &    10    & 42 &  8 & 930 \\
Raw data rate, $16nR$ [kbit s$^{-1}$] 
  & 220 & 845 & 264 & 352 & 560 & 1494 & 1443 \\[4pt]
\hline &&&&&&& \\[-4pt]
Raw data rate per instrument [kbit s$^{-1}$]
  & \multicolumn{4}{c|}{1681} 
  & \multicolumn{3}{c}{3496} \\[4pt]
\hline & \multicolumn{7}{c}{} \\[-4pt]
Total raw data rate [kbit s$^{-1}$]
  & \multicolumn{7}{c}{$2\times1681+3496=6858$} \\[4pt]
\hline
\end{tabular}
\label{tab:telemetry}
\vspace{-0.5cm}
\end{center}
\end{table*}

A `worst-case' star density, corresponding to about $2.8\times10^6$
stars per square degree (about 19--20~mag in Baade's Window) has been
used in a detailed analysis of CCD performance. The total noise per
sample includes contributions from the CCD read-out noise at the
relevant read frequency,  analog-to-digital conversion noise, and the
video chain analog noise.

All CCDs used for both astrometric and photometric measurements are
operated in the drift-scan or TDI (time delay and integration) mode.
That is, charge packets are gradually built up while transferred from
pixel to pixel at the same rate as the optical image moves across the
detector.  Centroiding on the digitised output provides a measure of
the  position of the optical image relative to the electronic transfer
along the pixel columns.  Demands on the precision  of the mean image
position are strict: a standard error of 10~$\mu$as in the final
trigonometric parallax translates to a required centroiding precision
in the astrometric field of 36~nm, 1/250 of a pixel. 

Specific laboratory experiments, using the $13\mu$m pixel EEV device CCD42-10
in windowing mode, non-irradiated as well as irradiated at doses of up
to $5\times10^9$~protons cm$^{-2}$, have been conducted in TDI mode,
using different illumination levels, and at different CCD operating
temperatures. Although not fully representative of the flight
configuration, and while not yet fully evaluated, these experiments 
have demonstrated that the targetted
centroiding accuracy appears to be achievable.

A key parameter for achieving a high degree of reproducibility
is the Charge Transfer Efficiency (CTE) of the CCD.  In the present
context it is more convenient to discuss the Charge Transfer 
Inefficiency (CTI) $\varepsilon=1-\mbox{CTE}$. Very few charge
carriers are actually lost (through recombination) during the 
transfer process; rather, some carriers are captured by `traps' and 
re-emitted at a later time, thus ending up in the `wrong' charge packet 
at the output; if short-time constant processes dominate, the main effect
observed is that of image smearing. The
CTI has an effect both on the photometric measurement (by reducing
the total charge remaining within the image) and the astrometric
measurement (by shifting charges systematically in one direction).
CTI during parallel transfer is
particularly critical, since it affects the astrometric measurements 
in the direction where the highest precision is required, i.e.\ along 
the scan. In addition, CTI is worse along-scan due to the lower 
transfer rate.
The magnitude of the problem can be crudely
estimated as follows: assume a constant fraction $\varepsilon$ of the
charge is left behind while the fraction $1-\varepsilon$ flows into
the next pixel.  The expected centroid shift is $\simeq
N\varepsilon/2$ pixels.  The CCDs in the astrometric field of GAIA
have $N=2780$ pixels of size 9~$\mu$m along the scan.  Assuming
$\varepsilon = 10^{-5}$ results in a centroid shift of 125~nm or about
500~$\mu$as. 

More careful appraisal of the CTI effects on the astrometric accuracy
show the effect after calibration is negligible for the the undamaged
(beginning-of-life) CCD, but potentially serious for the degraded
performance that may result after significant exposure to particle
radiation in orbit.  Although most of the CTI effects can be
calibrated as part of the normal data analysis, stochastic effects
related to the charge losses can never be eliminated by clever
processing. Extensive laboratory experiments are underway to quantify
the amplitude of these residual effects.

\begin{table}[t]
\caption{Summary properties of the GAIA CCDs in the two astrometric 
telescopes (QE~=~quantum efficiency; MTF~=~modulation transfer function; 
CTI~=~charge transfer inefficiency; RON~=~read-out noise).} 
\label{tab:ccd}
\begin{tabular}{ll}
\hline & \\[-5pt]
Feature&        Details \\[5pt]
\hline & \\[-5pt]
Array size&     $25\times58$~mm$^2$ active area \\
Pixels per CCD& 2150 cols $\times$ 2780 TDI stages \\
Dead zones&    top: 0.25~mm; sides: 0.6~mm; \\ 
&  bottom: $<5$~mm \\
Pixel size in image zone&  $9\times27~\mu$m$^2$ \\ 
Phases in image zone&           4 \\
Pixel size in serial register&  $27\times27~\mu$m$^2$ \\
Phases in serial register&      4 \\
Device thickness&    10--12~$\mu$m \\
Si resistivity&      20--100 $\Omega$cm \\
Buried channel&  n-type channel \\ 
Oxide thickness&    standard \\
Anti-blooming &   shielded at pixel level \\
Notch channel&   implanted for all CCDs \\
Output amplifiers&   2 per device, 2-stage \\
Conversion factor&  between 3--6~$\mu$V/e$^-$ \\
Additional gates&    5--10 \\
Power dissipation&    $<560$~mW \\
Non-uniformity&  $<1$\% rms local \\
	& $<10$\% peak-to-peak global \\ 
Mean dark current&  $<0.5$~e$^-$s$^{-1}$pix$^{-1}$ (200~K) \\
Non-linearity& $<1$~per cent over 0--2V; \\
 & $<20$~per cent over 2--3.5V \\
CTI in image area&   $<10^{-5}$ at beginning-of-life; \\
 &  $\sim10^{-4}$ after major Solar flare \\
CTI in serial register&  $<10^{-5}$ at beginning-of-life; \\
& $\sim5.10^{-4}$ after major flare \\
Quantum efficiency& trade with MTF and RON \\
MTF at Nyquist frequency& trade with QE and RON \\
Read-out noise& trade with QE and MTF \\
[5pt]
\hline
\end{tabular}
\end{table}

\subsection{Payload Summary}

In summary, the GAIA payload comprises the following elements:

\noindent (a) two identical astrometric telescopes:\\
\begin{tabular}{rl}
$-$ & fully-reflective 3-mirror SiC optics\\
$-$ & separation of viewing directions: 106$^\circ$\\
$-$ &  monolithic primary mirrors: $1.7\times0.7$~m$^2$\\
$-$ & field of view: 0.32~deg$^2$\\
$-$ & focal length: 50~m\\
$-$ & wavelength range: 300--1000~nm\\
$-$ & 4-colour broad-band photometry\\
$-$ & operating temperature: $\sim200$~K\\
$-$ & detectors: CCDs operating in TDI mode\\
$-$ & pixel size along-scan: 9~$\mu$m\\
\end{tabular}

\noindent (b) spectrometric instrument:\\
\begin{tabular}{rl}
$-$ & spectrometer for radial velocities\\
$-$ & 11 colour medium-band photometry\\
$-$ & entrance pupil: $0.75\times0.70$~m$^2$\\
$-$ & field of view: 4~deg$^2$\\
$-$ & focal length: 4.17~m \\
$-$ & detectors: CCDs operating in TDI mode\\
\end{tabular}

\section{Spacecraft System}

\subsection{Design and Operation}

The spacecraft subsystems provide all necessary support to the payload
instrumentation.  Designs follow well-established spacecraft
engineering approaches, with innovative features within the
mechanical, thermal, and telecommunication subsystems. Viable designs
have been developed for the mechanical structure, thermal control,
propulsion and attitude control, payload data handling, power and
electrical subsystems, and communications.

The payload mechanical and thermal design  provides the
required stability passively.  The stable orbit, combined with the
Sun shield thermal cover and the constant Solar aspect angle, minimise
external perturbations.  The payload material should have a low
coefficient of thermal expansion; high thermal conductivity to
variable heat loads; suitable structural and optical properties; and a
good light-weighting capability. Silicon carbide, as planned for
FIRST, Rosetta, and SOFIA, appears to be the optimum material.  The 
properties of silicon carbide allow it to be used both for the telescope
mirrors and the payload torus structure, providing a homogeneous, high
conductivity, athermal payload. The monolithic mirror sizes are 
compatible with present manufacturing capability. 

An open-back torus will support  the optical
instruments and focal plane, and electronic units, ensuring the
optical alignment is insensitive to uniform temperature variations.

The required line-of-sight stability (1~$\mu$as
rms) is high, while the short-term basic angle stability over
the satellite revolution period (3~hours) is the only critical
parameter so far identified which cannot be properly calibrated by
on-ground data processing. Detailed
thermal/mechanical analyses show that a basic angle variation of
1~$\mu$as rms corresponds to thermal gradient variations of
$\sim25~\mu$K in the payload module torus structure, and corresponding
displacements in the range of 20--40~pm. This motion need not be controlled,
but must be measured. A measurement device has been designed, and
proven in an industrial contract (TNO/TPD Delft). This real
measurement proved the viability of the basic angle measurement
technique, and  additionally demonstrated that a pattern on the CCD can be
localized with an accuracy compatible with the mission requirements.

The Sun shield is a simple, but large, multi-layer insulation disk, providing 
thermal stability for the payload module, integrated with the Solar panels.

The payload module opto-mechanical stability is sensitive to residual
thermal variations. The thermal control principle is based on
specific regulation and insulation for each thermal source, exploiting
the natural long time constant of the thermal cavity.
The service module thermal control design objectives are
to maintain electronic units inside the temperature specifications, 
and to avoid Sun reflections, hot spots or turning shadows on the
payload module cavity towards Sun shield and service module interface.
This is achieved, with reliance on the multi-layer insulation Sun shield.

\subsection{Attitude and Orbit Control}

The total $\Delta V$ budget required for injection into and
maintenance of the L2 operational orbit, including attitude control,
has been quantified. Transition between the transfer orbit and the
operational phase at L2 has been identified as one of the design
drivers of the attitude control subsystem.  The attitude control
measurement subsystem utilises a mix of star sensors, a Sun
acquisition sensor, payload instrument sky-mappers, a gyroscope (not
used in operational mode), an attitude anomaly detector (for safe
mode), bi-propellant thrusters, and FEEP (field-effect electric
propulsion) thrusters to control satellite motion. The solution with a
star sensor has been adopted since it provides simplification at
instrument level, at system level, and at mission level.  A detailed
evaluation of the relative merits of using spin-stabilisation or
3-axis stabilisation during the transfer phase has been carried out,
taking into account the overall system complexity, the required
thruster configuration, and the necessary propellant consumption. The
spin-stabilisation concept is adopted as baseline.

External perturbing sources are those induced by the Sun and the
gravitational effects of nearby planets. The latter effect concerns
only orbit drift. Solar pressures are created primarily by Solar
radiation and by the Solar wind. The Solar pressure is subject to
low-frequency variations which are modeled by an incoherent noise
superimposed on the (seismological) 5~minute oscillations.

The total spacecraft propellant budget is summarised as follows:
(i) orbit correction: 983~kg of
propellant are required; that is about one third of the launch mass, and
can be considerably reduced (to typically 200~kg) if the launcher
is able to directly inject the spacecraft on the transfer trajectory
to L2 (restartable Ariane~5);
(ii) transfer phase manoeuvres: 24~kg;
(iii) attitude control during operational phase: the Caesium
propellant budget for the FEEP thrusters in order to compensate for
the Sun radiation pressure disturbing torques is 0.1~kg over 6~years;
(iv) orbit maintenance during operational phase: 
1.6~kg of Caesium propellant.
In conclusion, the total propellant budget for the bi-propellant system is
1007~kg (including 10~per cent margin) and 2.7~kg for the FEEP-based system 
(including 50~per cent margin).

\subsection{Electrical, Power and Telemetry}

The spacecraft science data electrical architecture is complex, and
has several critical tasks.  There is one focal plane array per
astrometric instrument, one for the radial velocity spectrometer
and one for the medium-band photometer. Each focal plane
array includes CCD arrays and front end electronics.  The latter
includes the video preamplifiers as well as all the necessary bias and
voltage filters and clock drivers which have to be implemented close
to the detectors.  The focal plane array of each astrometric
instrument includes about 250 CCDs, corresponding to about 300 video
chains, has a mass of 40~kg, a power dissipation of 170~W, and
approximately 2500 interface cables with the video processing
unit. The detailed architecture of these focal plane arrays is a major
issue which requires further investigations.

The video processing units include all the video chains up to the
digitisation stage, the data discrimination function (star detection
and discrimination by mean of a fixed programmable threshold), the
localisation and datation of the detected events, the measurement of
the scan rate (data extracted from the sky mapper field) and the
transmission of these data to the spacecraft central computer for 
attitude control purposes, the multiplexing of the data and their 
transmission to the
data handling and processing unit. Assuming use of hybrids and ASICs
to improve the level of integration while minimising the power
dissipation, four video processing unit boxes per astrometric focal
plane are necessary for the hardware accommodation.
The other main items, the payload data handling unit, a 100~Gbit
solid-state recorder, and a high-rate telemetry formatter are 
not  critical items.

A centralised electrical architecture is proposed, in which 
a single computer (the central data management unit) provides all
the necessary control to all service module units, and
the required command and control signals to the payload electronics.
This minimisation of the quantity of hardware and software
corresponds to a general trend in the design of modern platforms, and
is realistic given the only moderate complexity of the thermal
control and attitude and pointing control functions.
This system, which is compatible with extant equipment, integrates a
power subsystem, thermal control hardware, the reaction control subsystem, 
the attitude control measurement subsystem, and the communications subsystems.

The preliminary power budget has been computed in observational mode and
with the spacecraft within ground station visibility, with
all instruments and payload units switched on, and with the science
telemetry down-link subsystem transmitting data to the ground
station. The bottom-line requirement is 641~W service module, 1527~W
payload, and 300~W contingency, for a total of 2468~W.
Adding interface losses to the spacecraft requirements, and
considering aging effects, a total Ga-As Solar array surface of 
24.1~m$^2$ is required.

A feasible technical solution has been identified for the
communication subsystem which is able to transmit the few Mbps
required for the science data based on a single ground station concept.
The Perth ground station presently offers the best compromise between
coverage and performance and is considered as the baseline. There is
no critical area identified at this stage for the telemetry and
telecommand link.  Omni-directional coverage will be provided by the
on-board antennae, to cope with any spacecraft attitude at any time
from launch through to the end of the mission. Relatively high RF
transmitted power is required (17~W) in order to provide the required
recovery margins from the L2 orbit.  Following the standard approach
for high-rate telemetry subsystems, this will require an additional
solid-state power amplifier at the transponder output.  Several
technology candidates for the science telemetry link antenna have been
identified and compared during the study: single beam fixed antenna,
switched antenna network, electronically scanned phased array antenna,
and single beam steerable antenna.  Detailed studies have led to the
adoption of an electronically scanned phased array (or `conformal')
antenna as baseline, with a single beam steerable antenna as back-up
solution.  The phased array has no moving parts, which means that
there is no source of dynamic perturbation for the attitude control
and measurement subsystem.  Although based on off-the-shelf
technologies at elementary level, it will require significant
engineering and development activities at assembly level, dedicated
design and arrangement of the radiating elements, and the overall
mechanical and thermal design.

\subsection{Orbit, Operations, and Ground Segment}

Detailed assesment of orbit options indicate that a Lissajous orbit
around the Earth-Sun Lagrange point L2 is the preferred option.  The
L2 region provides a very stable thermal environment (in order to
satisfy the stringent geometrical stability requirements of the
optical payload); an absence or minimisation of eclipses (which would
perturb the thermal environment and geometrical stability of the
payload, and thus require a more complex power subsystem and more
complex satellite operation); an absence of Earth or Moon occultations
(which would introduce straylight, thermal fluctuations, or blooming
at detector level); more stable perturbing torques (dominated by Solar
radiation pressure); and a lower radiation environment. The orbit is
compatible with an Ariane~5 dual launch, and results in the lowest
cost impact at system level, i.e.\ taking into account launch cost,
induced satellite complexity, and operations cost. The orbit is
consistent with a 6-year extended lifetime, with minimum mission
outages due to the transfer phase, perturbations, eclipses, etc.

The launch strategy selected as baseline is based on a dual or
multiple launch with Ariane~5, followed by injection of the satellite
from the standard geostationary transfer orbit into the L2~transfer
orbit via an autonomous propulsion system.  An alternative strategy
using a `restartable' Ariane~5, as foreseen for Rosetta and for
FIRST/Planck, would allow direct injection into the transfer orbit to
L2, and substantially reduce launch mass.

Orbits around the co-linear libration point L2 ($1.5\times10^6$~km
from the Earth away from the Sun) are planned for NGST,
FIRST/Planck, and GAIA.  Detailed analysis has selected for GAIA a
family of orbits around L2 which seen from the Earth describe a
Lissajous figure, have a time span from eclipse to eclipse of about
6~years, require minimal injection $\Delta V$, and retain the
Sun-spacecraft-Earth angle below $15^\circ$.

The astrometric analysis requires {\it a priori\/} knowledge of the
position and velocity vector of the satellite with respect to the
Solar System barycentre.  For positional precision, in the most
demanding case, taking $\pi=1$~arcsec and $\sigma_\pi=10$~microarcsec,
the satellite-barycentre distance needs to be known at the level of
$10^{-6}$~AU =~150~km. The Earth orbit is expected to be known with a
relative accuracy of $10^{-10}$ in the near future. For velocity
precision, the size of the correction is of the order of $v/c$, where
$v$ is the barycentric velocity of the observer and $c$ the velocity
of light.  In order to compute the correction to within
$\delta\alpha=1$~microarcsec ($5\times10^{-12}$~rad) the satellite
barycentric velocity vector is needed to within
$c\,\delta\alpha\sim1.5$~mm~s$^{-1}$. A prediction accuracy somewhat
better than 0.1~mm~s$^{-1}$ is realistic within a few years.

The radiation environment presents a hazard to space systems, in 
addition to the effects on the focal plane detectors. Penetrating
particles can induce upsets to electronics, payload interference,
damage to components and deep dielectric charging. 
Outside of possible Solar flares, L2~predictions of cosmic ray
particles have been derived using the well-established (CREME96) models.
ISO and HST experience is being applied in shielding specification,
although the total HST cosmic ray count rates are higher than for
GAIA, which at L2 is well out of Earth's magnetosphere.

\subsection{Spacecraft Summary}

The combined payload and service module results in a current
total mass of 1696~kg. Including a system margin of 20~per cent
results in a spacecraft dry mass of 2035~kg. The total satellite 
mass depends significantly on whether a liquid apogee engine 
is required (total launch mass 3137~kg), or whether the projected 
re-startable Ariane~5 capability allows this to be omitted (total launch 
mass 2267~kg).

In summary, the spacecraft and orbit are characterised as follows:

\begin{itemize}
\itemsep=0pt
\item orbit: Lissajous-type, eclipse-free, around L2 point of 
Sun/Earth system; 220--240~day transfer orbit 

\item sky scanning: revolving scanning with scan rate = 120~arcsec~s$^{-1}$,
precession period = 76~days

\item spacecraft: 3-axis stabilized; autonomous propulsion system for 
transfer orbit; electrical (FEEP) thrusters for operational attitude 
control; 6~deployable Solar panels, integrated with multi-layer 
insulation to form the Sun shield

\item science data rate: 1~Mbps sustained, 3~Mbps on down-link, using 
electronically steerable high-gain phased array antenna

\item launch mass: 3137~kg (payload = 803~kg, service module = 893~kg, 
system margin (20\%) = 339~kg, fuel = 1010~kg, launch adaptor = 92~kg)

\item power: 2569~W (payload = 1528~W, service module = 641~W, harness 
losses = 76~W, contingency (10\%) = 224~W)

\item payload dimensions: diameter = 4.2~m, height = 2.1~m

\item service module dimensions: diameter = 4.2~m (stowed)/8.5~m (deployed), 
height = 0.8~m

\item launcher: Ariane~5, dual launch 

\item lifetime: 5~years design lifetime (4~years observation time);
	 6~years extended lifetime
\end{itemize}

\section{Accuracy Assessment}

One important objective of the design studies has been to assess the overall 
performance of GAIA in
relation to its scientific goals.  This included the astrometric,
photometric and radial-velocity accuracies; the numbers of stars
observed to given accuracy levels; the diagnostic powers of the
resulting data; and limitations arising from the complexities of the
real sky.  

\subsection{Astrometric Accuracy}

There are three main components involved in the improved performance 
of GAIA compared with Hipparcos. The larger optics provide
a smaller diffraction pattern and a significantly larger collecting area;
the improved quantum
efficiency and bandwidth of the detector (CCD rather than
photocathode) leads to improved photon statistics; and use of CCDs
provides an important multiplexing advantage.

The GAIA astrometric wavelength band $G$ 
is fixed such that $G \simeq V$ for un-reddened A0V stars.
The approximate transformation from $(V,V\!-\!I)$ to $G$ 
can be expressed in the form:
\begin{eqnarray}
\label{equ:v-g}
  G = V + 0.51 &-& 0.50\times\sqrt{0.6+(V\!-\!I-0.6)^2} \nonumber \\
               &-& 0.065\times(V\!-\!I-0.6)^2
\end{eqnarray}
which is valid (to $\pm 0.1$~mag) at least for $-0.4<V\!-\!I\la 6$.
For $-0.4<V\!-\!I<1.4$ we have the convenient relation 
$G\!-\!V = 0.0 \pm 0.1$~mag. 

The sky background measured 
with HST at high ecliptic latitudes is $V=23.3$~mag~arcsec$^{-1}$.
Photometry and radial velocities calculations assume
$V=22.5$~mag~arcsec$^{-2}$. 

Many models of the Galaxy are available, from simplified star-count
models to complex evolutionary models. For present purposes we adopt a
star-count and kinematic model, updated from that of \cite*{che97}. This is
probably satisfactory within 25\%, and is adequate until GAIA
provides better data. The biggest uncertainty is perhaps the
distribution of interstellar extinction.
The model predicts that $\sim1.1$~billion stars will be observable,
corresponding to  1--2~per cent of the total stellar content of the
Galaxy. The estimated number of stars according to
the three Galactic populations (disk, thick disk and spheroid) shows
that  the main scientific goal, a representative census of the Galaxy,
can be achieved. In addition, there are many rare 
objects of high astrophysical interest for which star count
predictions cannot be obtained reliably from a Galaxy model.
Details include globular clusters, double and multiple stars,
high-density areas, and galaxies.

The basic accuracy estimation proceeds from the details of image 
formation, taking into account the detector signal, precision of the 
location estimator, instrument stability and calibration errors, and 
propagation from one-dimensional (single-epoch) measurement errors
to the final astrometric accuracy, estimated as:
\begin{equation} \label{equ:sigma_a}
  \sigma_a = g_a\left[\frac{\tau_1}{N_i\tau p_{\rm det}(G)} 
             (\sigma_\xi^2+\sigma_{\rm cal}^2)\right]^{1/2}
\end{equation}
where $N_i=2$ is the number of instruments, 
$\tau = L\Omega/4\pi$ the (average) total time available per object 
and instrument,
$\tau_1$ is the integration time per CCD (s),
$L$ is the effective mission length (s),
$\sigma_\xi$ is the angular precision in the scan direction from one 
CCD crossing (rad),
$\sigma_{\rm cal}$ is the accuracy of the astrometric calibration,
and  $p_{\rm det}(G)$ the detection probability as function of
magnitude.  
The factor $g_a$ relates the scanning geometry to the 
determination of the astrometric parameters.
Numerical simulations of
the scanning law are used to determine the mean values of $g_a$
for given mission parameters, and their large-scale variations
with ecliptic latitude.
The estimation of $p_{\rm det}$ is a complex problem, since it
depends on many factors besides the brightness of the star.  For
the accuracy analysis an estimation of $p_{\rm det}$ as function
of $N$, the total number of photoelectrons in the stellar image,
was made by means of dedicated simulations. 
An explicit error margin of 20~per cent is added on the astrometric
standard errors resulting from this analysis.

For the faintest stars the accuracy estimates take into account 
that a given star may not be observed in all transits.  They 
include the factor $p_{\rm det}(G)^{-1/2}$, where $p_{\rm det}(G)$ 
is the detection probability as a function of the $G$ magnitude 
obtained from simulating star detection in the astrometric sky mapper
using the robust APM algorithm. Since, by construction, the $G$ magnitude
yields a rather uniform accuracy as a function of spectral type,
useful mean accuracies can be derived from a straight mean of the
detailed calculations, and are given in Table~\ref{tab:accuracy}.

An approximate analytical fit to the tabular values for the 
parallax accuracy, $\sigma_\pi$, which also takes 
into account the slight colour dependence (due to the widening of the 
point spread function at longer wavelengths) is:
\begin{eqnarray}
  \label{equ:approx-acc}
  \sigma_\pi \simeq && (7+105\,z+1.3\,z^2+6\times 10^{-10}\,z^6)^{1/2} 
	\times \nonumber \\
             && \qquad\qquad \left[ 0.96 + 0.04\,(V\!-\!I) \right]
\end{eqnarray}
where $z = 10^{0.4(G-15)}$.  For the position and proper motions errors, 
$\sigma_0$ and $\sigma_\mu$, the following mean relations can be used: 
\begin{eqnarray} 
  \sigma_0 &=& 0.87\,\sigma_\pi \\
  \sigma_\mu &=& 0.75\,\sigma_\pi
\end{eqnarray}
The expected standard errors vary somewhat over the sky
as a result of the scanning law.

\begin{table*}
\caption{Mean accuracy in parallax ($\sigma_\pi$), position 
(at mid-epoch, $\sigma_0$) and proper motion ($\sigma_\mu$), 
versus $G$ magnitude.  The values are sky averages.}
\label{tab:accuracy}
\vspace{-5pt}
\begin{center}
\footnotesize
\begin{tabular}{lrrrrrrrrrrrr}
\hline &&&&&&&&&&&& \\[-5pt]
$G$~(mag)\quad&    10& 11& 12& 13& 14& 15& 16& 17& 18& 19&  20& 21\\[4pt]
\hline &&&&&&&&&&&& \\[-5pt]
$\sigma_\pi$~($\mu$as)& 4& 4& 4& 5& 7& 11& 17& 27& 45& 80& 160& 500\\
$\sigma_0$~($\mu$as)& 3& 3& 3& 4& 6& 9& 15& 23& 39& 70& 140& 440\\
$\sigma_\mu$~($\mu$as~yr$^{-1}$)&3& 3& 3& 4& 5& 8& 13& 20& 34& 60& 120& 380\\[4pt]
\hline
\end{tabular}
\end{center}
\end{table*}

\begin{figure*}[t]
\begin{center}
\leavevmode
\centerline{
\epsfig{file=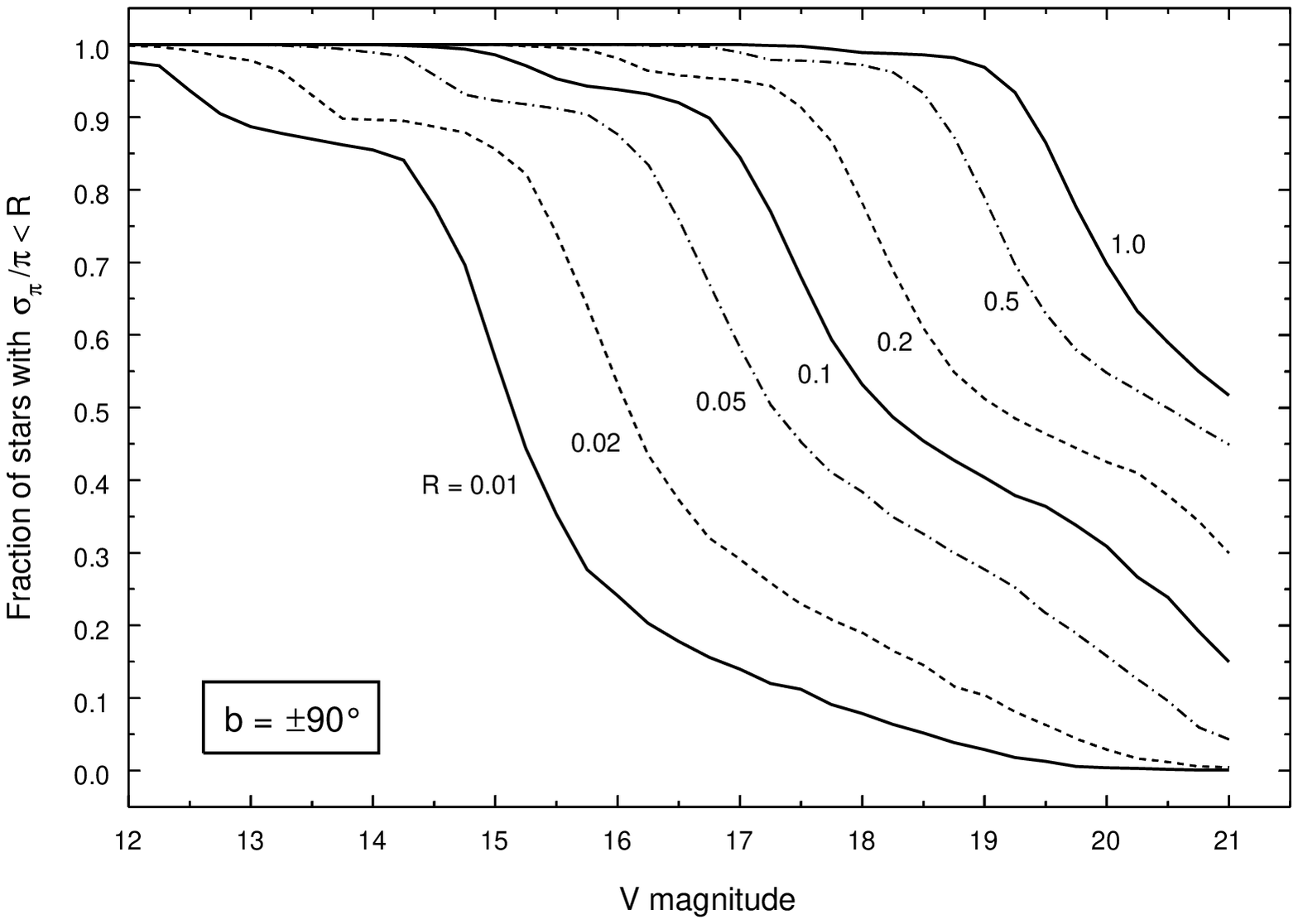,width=8.8cm,angle=0}
\hfill
\epsfig{file=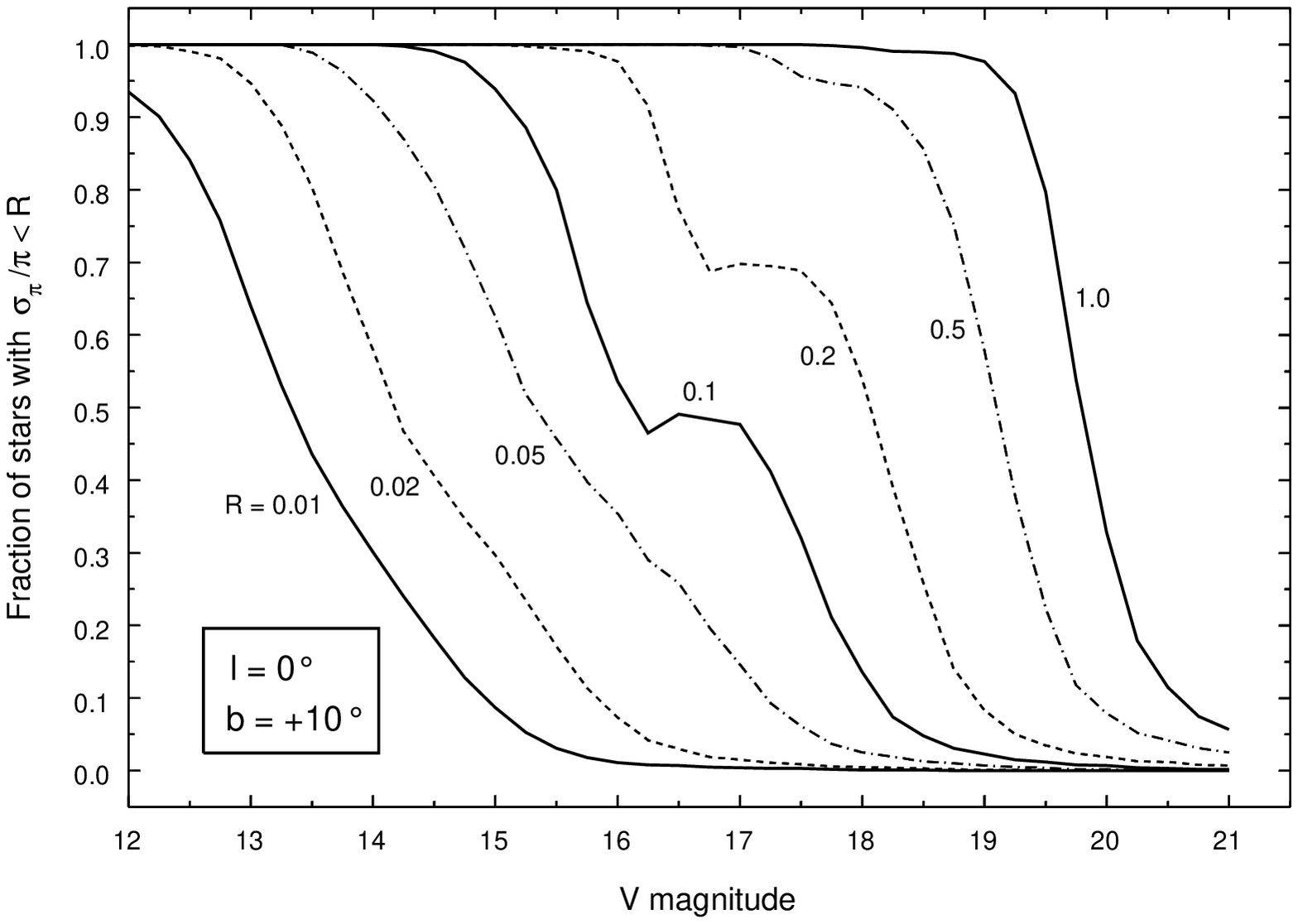,width=8.8cm,angle=0} 
}
\end{center}
\vskip -10pt
\caption{Left: fraction of stars at a given magnitude having 
a relative parallax error less than 1, 2, 5, 10, 20, 50 and 100~per cent, 
along the line of sight towards $(l=0^\circ,b=+90^\circ)$.
Right: the same, for Galactic coordinates $(l=0^\circ,b=+10^\circ)$.}
\label{fig:accuracy}
\end{figure*}

The number of stars whose distances can be determined to a certain
relative accuracy can be estimated using a Galaxy model. For a given
apparent magnitude and direction, the model provides the distribution
of stars along the line of sight (in a small solid angle), as well as their
distribution in colour index at each distance.  It is then simple to
compute the fraction of stars having relative parallax accuracy below
a certain limit $R$.  Figure~\ref{fig:accuracy} shows, for selected
directions, the fraction of stars with relative parallax accuracies
below $R=0.01-1.0$.  Very good distance information ($R=0.1$ or
better) will be obtained for virtually all stars brighter than $V=15$
and for significant fractions down to much fainter magnitudes, e.g.\
10--50~per cent at $V \sim 18$, depending on direction. 

For $G\leq15$~mag, more than 85~per cent of the stars will have 
tangential velocity errors
smaller than 5~km~s$^{-1}$, and 75~per cent will be
smaller than 2~km~s$^{-1}$. These figures worsen if all the stars with
$G\leq20$~mag are considered, but even in this case, for galactic
latitudes higher than $5-10^\circ$, 40~per cent of the stars
observed by GAIA will have tangential velocities accurate to better than
10~km~s$^{-1}$.

\subsection{Photometric Accuracy}

The photometric analysis cannot be separated from the
astrometric analysis.  The basic model for the pixel or sample
values must be fitted simultaneously for the centroid coordinates
$\xi_0$, $\eta_0$ and the photometric quantities $b$ and $N$.
Moreover, this fitting must be done globally, by considering
together all the transits of the object throughout the mission,
in which the centroid positions are constrained by the 
appropriate astrometric model.  This procedure is referred to
as `global PSF fitting'.
A simpler procedure,  `aperture photometry',
is used for the photometric accuracy estimation.

Simple calculations for single-epoch accuracies 
including photon noise, read-noise and 
sky background noise have been adopted, and corrected with 
an error margin of 20~per cent. 
The formal photometric {\it precision\/} reached at the end of the
mission, typically by averaging $n_{\rm obs}\sim
50$--$100$~observations, is on the level of one or a few millimag for
the bright stars ($\sim 12$~mag). Can this be calibrated to reach this
{\it accuracy}?  Photometric calibration of the CCD zero points can be
achieved from standards. The photometric consistency can be derived
from repeated observations of all non-variable stars, and assured to a
very high degree.

Photometric observations in a number of colour bands 
will be obtained for all stars detected.  A quick
photometric reduction will be carried out for each field crossing,
providing rapid epoch photometry results for 
scientifically time-critical phenomena
such as the detection of supernovae, burst, lensing, or other
transient events.  Improved photometric reduction can
be obtained later in the mission, when accurate satellite attitude,
CCD calibration data, and astrometric information for each star become
available.  This accuracy has been simulated using the complete GAIA
image simulation and the APM photometry package.  The two astrometric telescopes
provide photometry in a wide spectral band defined by the CCD
sensitivity. This $G$ band photometry from one single measure has a
standard error of typically 0.01~mag for $G\sim18.5$~mag.  The two
astrometric telescopes also provide broad-band photometry in four bands.
The magnitudes
resulting from averaging 67~observations obtained during a mission
time of 4~years will have a precision of about 0.02~mag in the F63B
band (see Figure~\ref{fig:photometric-system}) at $V\sim20.0$~mag for 
all spectral types.  The spectrometric telescope will collect photometry 
in 11~bands. The resulting average magnitudes will have a precision
of 0.01~mag in the F57 band at $V=19$~mag, and 0.02~mag in the
F33~band for an unreddened G2V star.

\begin{figure*}[t]
\begin{center}                                                       
\leavevmode                                                           
\centerline{
\epsfig{file=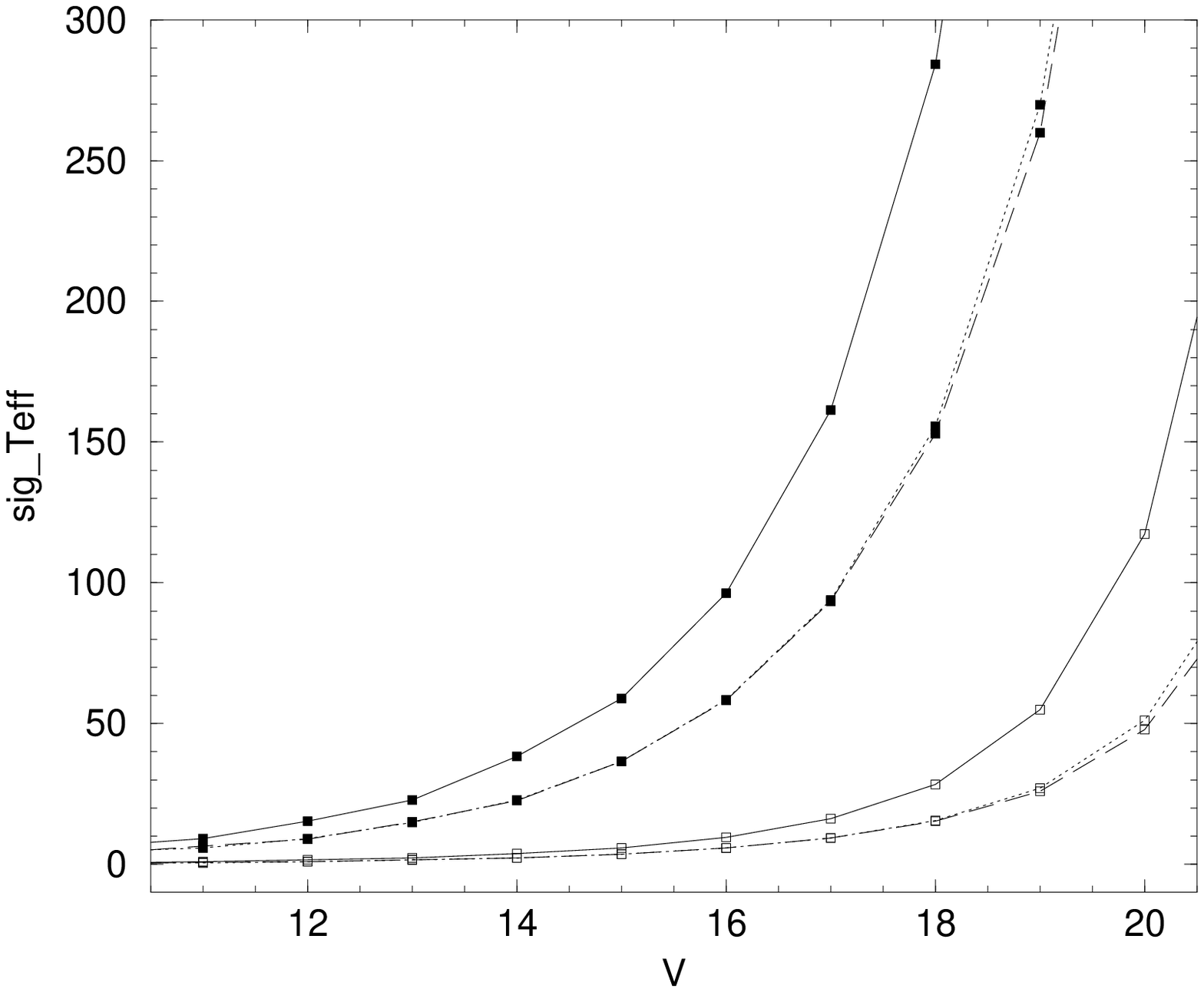,width=8.8cm,angle=0}   
\hfill                                                          
\epsfig{file=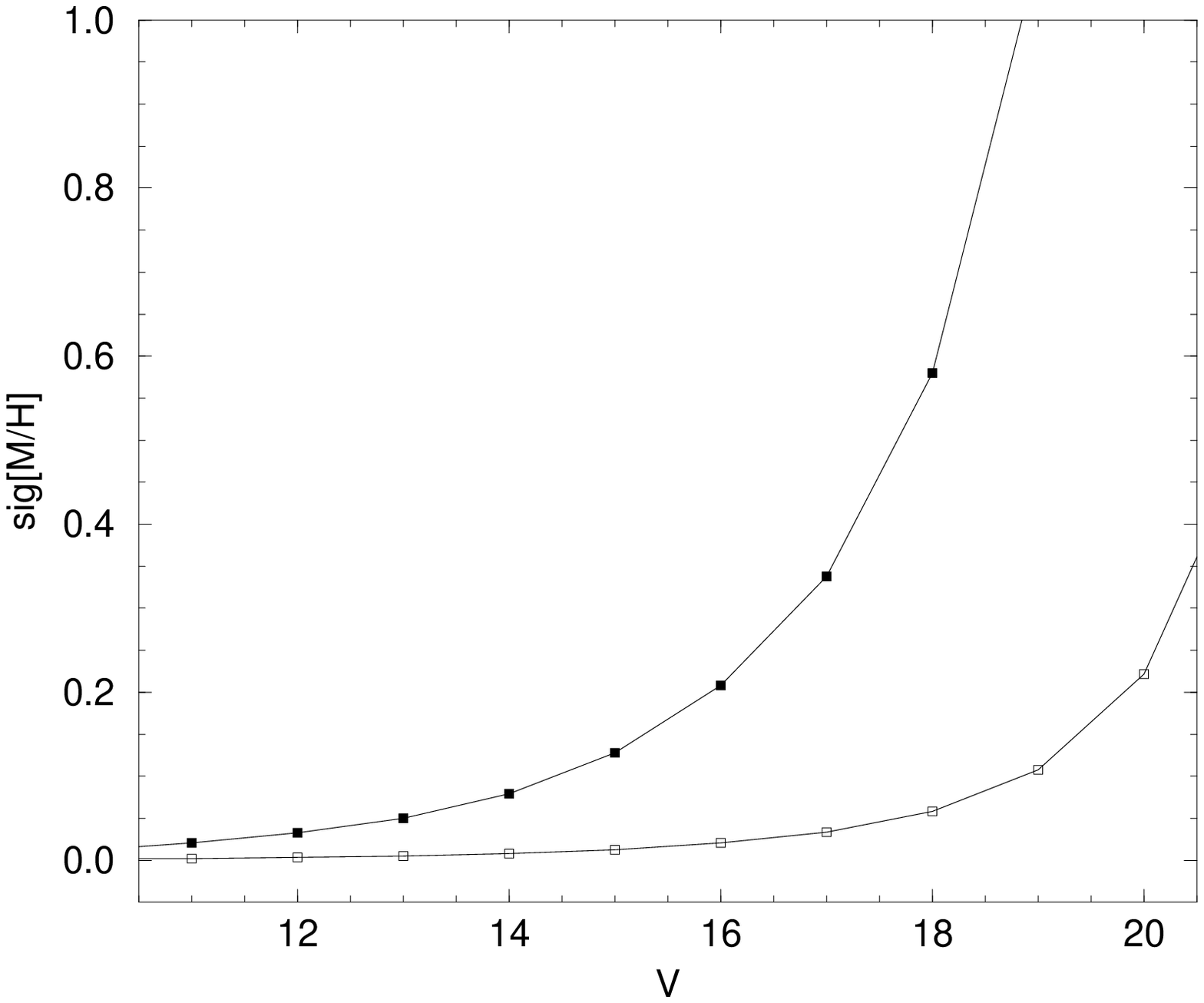,width=8.8cm,angle=0}
} 
\end{center}                                                       
\vskip -10pt                                           
\caption{Left: errors on $T_{\rm eff}$ derived from C47--57
(continuous line), C57--75 (dotted line) and C75--89 (dashed line)  
indices, for M dwarfs with $T_{\rm eff}$~=~3500~K.        
Solid symbols correspond to                                      
the errors for a single observation, open symbols to the             
mission-average (100~observations).
Right: errors on [Ti/H] derived from the TiO index, for
$T_{\rm eff}$~=~3500 K M dwarfs,
for single observations (filled symbols) and for
mission averages (open symbols).}
\label{fig:accuracy-diagnostics}
\end{figure*}

The photometric accuracies as a function of colour index 
can be converted into accuracies on the astrophysical
parameters. As an example, Figure~\ref{fig:accuracy-diagnostics}(left)
shows the uncertainty in effective temperature for G- and M-type
dwarfs ($T_{\rm eff}$~=~5750 and 3500~K respectively) as a function of
magnitude, for single transit and mission-average photometry, while
Figure~\ref{fig:accuracy-diagnostics}(right) shows the uncertainty in
[Ti/H] for M~dwarfs.

\subsection{Radial Velocity Accuracy} 

The radial velocity accuracy assessment has been derived from 
extensive numerical simulations, utilising observed spectra.
Template spectra covering a wide range of astrophysical properties
will be required. Global velocity zero points can be derived from the
system geometry and astrometric positions of the target stars. 
The required accuracy is 1/40 pixel for 1~km~s$^{-1}$. 
Performance has been studied using simulations as well as real observations 
of the selected spectral region for a variety of stars.  Simulations were
performed by producing synthetic spectra with different
atmospheric parameters in the Ca\,{\sc ii} region, degrading
them to the resolution and signal-to-noise ratio of a single GAIA
observation, and then determining the radial velocity by 
cross-correlation. The photon budget was computed from the
basic instrument design, and in addition to Poisson noise both
the total read-out noise (3~e$^-$) and the sky background (normalized
to a Solar spectrum with $V=22.5$~arcsec$^{-2}$) were considered.
For cool stars, the mission-average
velocity accuracy is $\sigma_v \simeq 5$~km~s$^{-1}$ at $V=18$, while
for hot stars the performance is limited to $\sigma_v \simeq
10$~km~s$^{-1}$ at $V=16$~mag.

\subsection{Summary}

In summary, GAIA's measurement capabilities can be summarised as 
follows:

Catalogue: $\sim1$~billion stars;
$0.34\times10^6$ to $V=10$~mag;
$26\times10^6$ to $V=15$~mag;
$250\times10^6$ to $V=18$~mag;
$1000\times10^6$ to $V=20$~mag; 
completeness to about 20~mag. 

Sky density:
mean density $\sim25\,000$ stars deg$^{-2}$;
maximum density $\sim3\times10^6$ stars deg$^{-2}$. 

Median parallax errors:
4~$\mu$as at 10~mag;
11~$\mu$as at 15~mag;
160~$\mu$as at 20~mag.

Distance accuracies: 
2~million better than 1~per cent;
50~million better than 2~per cent; 
110~million better than 5~per cent; 
220~million better than 10~per cent.

Tangential velocity accuracies:
40~million better than  0.5~km~s$^{-1}$;
80~million better than  1~km~s$^{-1}$;
200~million better than  3~km~s$^{-1}$;
300~million better than  5~km~s$^{-1}$;
440~million better than 10~km~s$^{-1}$. 

Radial velocity accuracies: 1--10~km s$^{-1}$ to $V=16-17$~mag,
depending on spectral type.

Photometry: to $V=20$~mag in 4~broad and 11~medium bands.

\section{Data Analysis}

\begin{figure*}[t]
\begin{center}
\leavevmode
\centerline{\epsfig{file=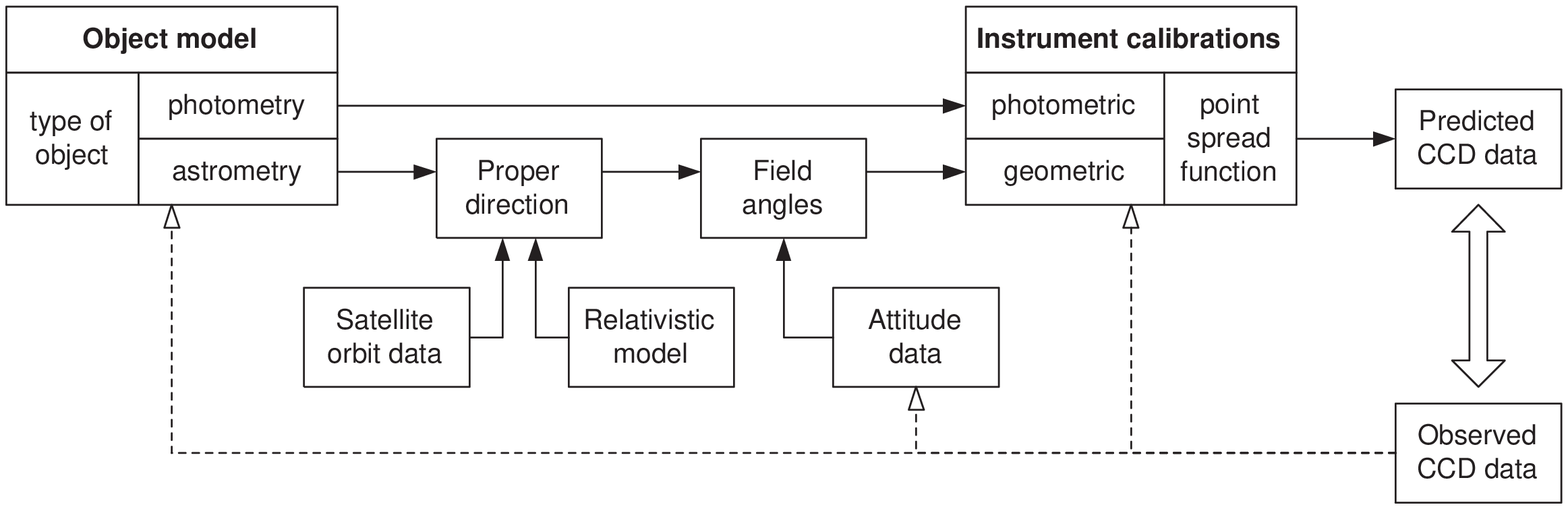,width=16cm,angle=0}}
\end{center}
\vskip -10pt
\caption{Model of CCD data interrelations for an astronomical object.
In principle, the data analysis aims to provide the `best' representation
of the observed data in terms of the object model, satellite attitude and
instrument calibration.  Certain data and models can, from the viewpoint
of the data analysis, be regarded as `given'; in the figure these are
represented by the satellite orbit (in the barycentric reference system) 
and the relativistic model used to compute celestial directions.  Other 
model data are adjusted to fit the observations (dashed lines).}
\label{fig:ll_da-flow}
\end{figure*}

The total amount of (compressed) science data generated in the course
of the five-year mission is about $2\times 10^{13}$~bytes (20~TB).  
Most of this consists of CCD raw or binned pixel values with associated
identification tags.  The data analysis aims to 
`explain' these values in terms of astronomical objects and their
characteristics. In principle the analysis is done by
adjusting the object, attitude and instrument models until a satisfactory
agreement is found between predicted and observed data (dashed lines
in Figure~\ref{fig:ll_da-flow}). Successful implementation of the 
data analysis task will require expert knowledge from several
different fields of astronomy, mathematics and computer science to be
merged in a single, highly efficient system (\cite{ml99}).

The global astrometric reductions must be formulated in a
fully general relativistic framework, including
post-post-Newtonian effects of the spherical Sun at the 1~$\mu$as
level, as well as including corrections due to oblateness and angular
momentum of Solar System bodies.

Processing these vast amounts of data will
require highly automated and efficient numerical methods.
This is particularly critical for the image centroiding of 
the elementary astrometric and photometric observation in the astrometric 
instruments, and the corresponding analysis of spectral data in the 
spectrometric instrument.  

Accurate and efficient estimation of the centroid coordinate based on the 
noisy CCD samples is crucial for the astrometric performance.  
Simulations indicate that 6~samples
approximately centred on the peak can be read out from the CCD.  The
centroiding, as well as the magnitude estimation, must be based on
these six values.  Results of a large number of Monte Carlo
experiments, using a maximum-likelihood estimator as the centroiding
algorithm, indicate that a rather simple
maximum-likelihood algorithm performs extremely well under these
idealized conditions, and that six samples is sufficient to determine
the centroid accurately.  Much work remains to extend the analysis to
more complex cases, including in particular overlapping stellar images.

A preliminary photometric analysis, for discovery of variables,
supernovae, etc, can be carried out using standard photometric
techniques immediately after data delivery to the ground. In addition,
more detailed modelling of the local background and structure in the
vicinity of each target using all the mission data in all the
passbands will be required. A final end-of-mission re-analysis may
benefit from the astrometric determination of the image centroids,
locating a well-calibrated point spread function for photometric analysis.
Studies of these photometric reductions have begun.

The high-resolution (radial velocity) spectrometer will produce spectra for 
about a hundred million stars, and multi-epoch, multi-band photometry will be
obtained for about one billion stars. The analysis of such large
numbers of spectra and photometric measurements needs to be performed
in a fully automated fashion, with no manual intervention. Automatic
determination of (at least) the surface temperature $T_{\rm eff}$, the
metallicity [M/H], and the relative $\alpha$ element abundance
[$\alpha$/Fe] is necessary; determination of $\log g$ is, given the
availability of parallaxes for most stars, of lesser importance.
A fully automated system for the derivation of
astrophysical parameters from the large number of spectra and
magnitudes collected by GAIA, using all the available information for
each star, has  been studied, showing the feasibility of an approach
based on the use of neural networks. 
In the classification system foreseen, spectra and photometric
measurements will be sent to an `initial classifier', to sort objects
into stellar and non-stellar. Specialist networks then treat each
class. For example, stellar data sets are passed to an `automated
stellar parameterization' sub-package.

It is the physical parameters of stars which are
really of interest; therefore the proposed system aims to derive
physical parameters directly from a stellar spectrum and photometry. 
Detailed simulations of the automated stellar parameterization
system have been completed using a feed-forward 
neural network operating on the entire set of spectral and photometric
measurements. In such a system,
the derived values for the stellar parameters are  naturally
linked to the models used to train the network. Given the extreme
rapidity of neural networks, when stellar atmosphere models are improved,
re-classification of the entire data set can be done extremely
quickly: an archive of $10^8$ spectra or photometric measurements
could be reclassified in about a day with the present-day computing
power of a scientific workstation.

The overall data analysis task would be impossible without certain regularizing 
assumptions: one must 
assume that a substantial fraction of stars follow a very simple model, 
viz.\ (apparently) single stars with little or no photometric 
variability, whose motions can be described by the standard five 
astrometric parameters ($\alpha$, $\delta$, $\pi$, $\mu_{\alpha*}$,
$\mu_\delta$).  For the satellite attitude and instrument 
characteristics it must be assumed that sudden changes are rare, 
so that time-averaging and smoothing are effective in reducing 
observational noise.  Without these assumptions the problem would
simply have too many degrees of freedom.
While such regularity conditions must be valid in a broad sense, it 
is clear that they cannot be guaranteed to hold in any 
particular situation or for a specific object.  The data analysis
must be able to filter out cases where the conditions do not
apply, and divert them to a separate analysis branch. The
efficiency of the filtering process depends critically on the 
quality of the instrument calibrations and attitude determination, 
which initially is quite low.  Thus an iterative process is needed
in which the object selection and observations are successively
improved, along with the calibrations and attitude determination.

The computational complexity of the data analysis arises not just 
from the amount of data to be processed, but even more from the 
intricate relationships between the different pieces of information 
gathered by the various instruments throughout the mission.
It is difficult to assess the magnitude of the data analysis
problem in terms of processing requirements.  Certain basic algorithms
that have to be applied to large data sets can be translated into a
minimum required number of floating-point operations. Various
estimates suggest of order $10^{19}$~floating-point operations,
indicating that very serious attention must be
given to the implementation of the data analysis, and that this effort must
start very early.

Observations of each object are distributed throughout the mission, 
so that calibrations and analysis must be feasible both in the 
time-domain and in the object domain. Flexibility and
interaction is needed to cope with special objects, while
calibrations
must be protected from unintentional modification. 
Object Oriented (OO) methodologies for data modeling, storage and
processing are ideal for meeting the challenges faced by GAIA.

The feasibility of the OO approach has been demonstrated by a short
prototyping exercise carried out during the present study phase.
Algorithms for three processes were provided and incorporated into the
OO model, underlining one important feature of OO design: the ability
to have complex data structures and operations described in a single
model. Java code was generated from the model and the algorithms
implemented.  The prototype was highly successful and reinforced
confidence in the OO approach for treating the data.  
The reduction process is inherently distributed, and
naturally matched to distributed parallel processors. 

\section{GAIA and other Space Missions}
\label{sec:other-missions}

The scientific capabilities and goals of GAIA and other proposed or
approved space astrometric missions are summarised in
Table~\ref{tab:missions}. GAIA is a survey mission, essential for
statistical analysis of the unknown, with broad applications to the
Solar System, galaxies, large-scale structure, and primarily Galactic
structure and evolution. SIM (Space Interferometry Mission,
\cite{bus97}) is an interferometer, ideal for precise measurements of
a small number of carefully pre-selected targets of specific
scientific interest. The SIM target selection is yet to happen, but
will be focussed towards searches for low-mass planets around a few
nearby stars, calibration of the distance scale, and detailed studies
of known micro-lensing events.

FAME (Full-Sky Astrometric Mapping Explorer; \cite{tgh+00}), selected
by NASA in the MIDEX competition in 1999, and DIVA (Double 
Interferometer for Visual Astrometry; 
\cite{rbb+97}), a small German astronomy satellite planned for launch
in 2004, are essentially successors to Hipparcos, with an extension of
limiting sensitivity, sample size and accuracy (statistical weight) by
a factor of order 100 in each. DIVA and FAME will both provide an
excellent reference frame, substantially improve calibration of the
distance scale and the main phases of stellar evolutionary
astrophysics, and map the Solar Neighbourhood to much improved
precision. The difference between GAIA and these two missions is one
of scale and comprehensiveness. GAIA exceeds these two missions in
scale by a further factor of order 100, allowing study of the entire
Galaxy, and only GAIA will provide photometric and radial velocity
measurements as crucial astrophysical diagnostics.

\begin{table}[t]
\caption{Summary of the capabilities of Hipparcos and GAIA, along with 
those of the DIVA (Germany) and SIM and FAME (NASA) 
astrometric space missions. Numbers of stars are 
indicative; in the case of SIM they are distributed amongst grid stars
and more general scientific targets. Typical accuracies are given 
according to magnitude where appropriate.}
\label{tab:missions}
\vspace{1pt}
\begin{center}
\leavevmode
\footnotesize
\begin{tabular}{lcrccc}
\hline &&&&& \\[-5pt]
Mission&    Launch& No. of&		Mag&	\multicolumn{2}{c}{Accuracy} \\
&	    &	stars&		limit&	(mas)& (mag) \\[3pt]
\hline &&&&& \\[-5pt]
Hipparcos&  1989&	120\,000&	12&	1\phantom{.000}& 10 \\[3pt]
DIVA&	    2004&	40 million&	15&	0.2\phantom{00}& \phantom{0}9 \\
	&   	&	&		&	5\phantom{.000}& 15 \\[3pt]
FAME&	    2004&	40 million&	15&	0.050& \phantom{0}9 \\
	&   	&	&		&	0.300& 15 \\[3pt]
SIM&	    2009&	$10\,000$&	20&	0.003& 20 \\[3pt]
GAIA&	    2012&	$1$~billion&	20&	0.003& 12 \\
	&   	&	&		&	0.010& 15 \\
	&   	&	&		&	0.200& 20 \\[3pt]
\hline
\end{tabular}
\end{center}
\end{table}

\section{Conclusion}

GAIA will create an extraordinarily precise three-dimensional map of
about one billion stars throughout our Galaxy and the Local group. It
will map their space motions, which encode the origin and subsequent
evolution of the Galaxy, and the distribution of dark matter.  Through
on-board photometry, it will provide the detailed physical properties
of each star observed: luminosity, temperature, gravity, and elemental
composition, which encode the star formation and chemical enrichment
history of the Galaxy. Radial velocity measurements on board will
complete the kinematic information for a significant fraction of the
objects observed. 

Through continuous sky scanning, the satellite will repeatedly measure
positions and colours of all objects down to $V=20$~mag. On-board
object detection ensures a complete census, including variable stars
and quasars, supernovae, and minor planets. It also circumvents costly
pre-launch target selection activities. Final accuracies of
10~microarcsec at 15~mag will provide distances accurate to 10~per
cent as far as the Galactic Centre. Stellar motions will be measured
even in the Andromeda galaxy.

In order to limit failure modes, the instrument design includes only 
one deployable element (the sun shield/solar array) and only three
on-board mechanisms (two secondary mirror correctors, one for each of 
the astrometric telescopes, and an orientation
adjustment for the spectroscopic CCDs). Operation of the focal 
plane is robust against failure of individual CCDs, or against failure of 
one or more `rows' of the main focal plane (whether it be the sky mapper, 
the main detectors, or the on-board data handling units associated with each 
such row), which would correspond to no more than a loss of overall 
observing time. The CCDs for the medium-band photometer and radial 
velocity spectrograph are designed so that the loss if one CCD 
leads to the loss of only the upper or lower half of one colour band,
amounting to a `graceful degradation' of the mission's science goals.
Loss of throughput, of whatever form, would lead to a decrease in the
limiting magnitude, and a corresponding degradation of the astrometric
accuracy as a function of magnitude. It is not easy to quantify `break
points' in the scientific case at which the astrometric improvement
compared to Hipparcos, occasioned by failures or performance
limitations, would cease to have a significant scientific impact, as
evidenced by the selection of both the FAME and DIVA missions.
Nonetheless, a limiting magnitude of 20~mag, accuracies of
10~$\mu$arcsec at 15~mag, and the provision of in-depth photometric
and radial velocity data for each object, remain primary mission goals.

GAIA's main scientific objective is to clarify the origin and history of our 
Galaxy, from a quantitative census of the stellar populations. It will
advance fundamental questions such as when the stars in the Milky Way 
formed, when and how the Milky Way was assembled, and the 
distribution of dark matter in our Galaxy. In so doing, it
will pinpoint exotic objects in substantial numbers: many thousands of
extra-Solar planets will be discovered, and their detailed orbits
determined; tens of thousands of brown dwarfs and white dwarfs will be
identified; rare stages of stellar evolution will be quantified; some
100\,000 extragalactic supernovae will be discovered, and details
communicated for follow-up ground-based observations; Solar System
studies will receive a massive impetus through the detection of many
tens of thousands of new minor planets; inner Trojans and even new
trans-Neptunian objects, including Plutinos, may be discovered. GAIA
will follow the bending of star light by the Sun and major planets
over the entire celestial sphere, and therefore directly observe the
structure of space-time---the accuracy of its measurement of general
relativistic light bending may reveal the long-sought scalar
correction to its tensor form.  The PPN parameters $\gamma$ and
$\beta$, and the Solar quadrupole moment J$_2$, will be determined
with unprecedented precision. New constraints on the rate of change of
the gravitational constant, $\dot{G}$, and on gravitational wave
energy over a certain frequency range, will be obtained.

GAIA is timely as it complements other major space and ground
initiatives. Understanding and exploration of the early Universe,
through microwave background studies (Planck) and direct observations
of high-redshift galaxies (NGST, FIRST, ALMA), are complemented by
theoretical advances in understanding the growth of structure from the
early Universe up to galaxy formation. Serious further advances
require a detailed understanding of a `typical' galaxy, to test the
physics and assumptions in the models. Our Galaxy, a typical example
of those luminous spirals which dominate the luminosity of the
Universe, uniquely provides such a template.

\begin{acknowledgements}
This summary of the GAIA mission, as presented to and approved by the
scientific advisory committees of the European Space Agency in
September--October 2000, is based on the GAIA Study Report, which is
the result of a large collaboration between ESA, the European scientic
community and European industry. The scientific aspects of the study
were supervised by the Science Advisory Group Members, comprised by
the authors. The work of the Science Advisory Group has been supported
by a Science Working Group, chaired by P.T.~de Zeeuw and G.~Gilmore
(18 members) and responsible for quantifying the science case; a
Photometry Working Group, chaired by F.~Favata (18 members); an
Instrument Working Group, chaired by L.~Lindegren (16 members); and 52
other European scientists directly supporting the GAIA study. We 
gratefully acknowledge the many and detailed contributions made by the 
working group members, as well as guidance on the content of this paper 
from the referee, Donald J.~Hutter.

Technological activities have been led by the ESA Study Manager,
O.~Pace, supported by M.~Hechler (ESOC Study Manager), and
ESA--ESTEC engineers. ESA scientific involvement includes
contributions from S.~Volont\'e (ESA Paris) and from scientists within
the ESA Astrophysics Division (K.S.~O'Flaherty, F.~Favata, W.~O'Mullane,
M.~Vannier, and A.~Colorado McEvoy).

The satellite design study has been under contract to Astrium (formerly 
Matra Marconi Space, F), under Study Manager P.~M\'erat, and involving EEV Ltd
(UK), and Alcatel Space (F). An Alenia Study Team, under Study Manager
S.~Cesare, evaluated the performance of an interferometric design,
with involvement of the Istituto di Metrologia `G.~Colonnetti', EICAS
Automazione, the Osservatorio Astronomico di Torino, Matra Marconi
Space (F), and Alcatel Space.  Other industrial studies have been
carried out by SIRA (UK; CCD CTE), and TNO--TPD (Delft; basic angle
monitoring).

\end{acknowledgements}

\bibliographystyle{aabib}

\begin{thebibliography}{}

\bibitem[\protect\astroncite{Battrick}{1994}]{bat94}
Battrick B. 1994,
\newblock in B. Battrick (ed.), Horizon 2000 Plus: European Space Science in
  the 21st Century, ESA SP-1180, ESA, Noordwijk

\bibitem[\protect\astroncite{Boden et~al.}{1997}]{bus97}
Boden A., Unwin S., Shao M. 1997,
\newblock in M.~A.~C. Perryman, P.~L. Bernacca (eds.), Hipparcos Venice 97, ESA
  SP-402, ESA, Noordwijk, p.~789

\bibitem[\protect\astroncite{Chen}{1997}]{che97}
Chen B. 1997, ApJ 491, 181

\bibitem[\protect\astroncite{de~Zeeuw}{1999}]{zee99}
de~Zeeuw P.~T. 1999,
\newblock in B.~K. Gibson, T.~S. Axelrod, M.~E. Putman (eds.), The Third
  Stromlo Symposium: The Galactic Halo: Bright Stars and Dark Matter, ASP Conf.
  Ser. 165, p.~515

\bibitem[\protect\astroncite{ESA}{2000}]{esa-2000-4}
ESA 2000,
\newblock GAIA: Composition, Formation and Evolution of the Galaxy,
\newblock Technical Report ESA-SCI(2000)4,
\newblock (scientific case on-line at http://astro.estec.esa.nl/GAIA)

\bibitem[\protect\astroncite{Eyer \& Cuypers}{2000}]{ec00}
Eyer L., Cuypers J. 2000,
\newblock in L. Szabados, D.~W. Kurtz (eds.), The Impact of Large-Scale Surveys
  on Pulsating Star Research, ASP Conf. Ser. 203, p.~71

\bibitem[\protect\astroncite{Feissel \& Mignard}{1998}]{fm98}
Feissel M., Mignard F. 1998, A\&A 331, L33

\bibitem[\protect\astroncite{Freeman}{1993}]{fre93}
Freeman K.~C. 1993,
\newblock in S. Majewksi (ed.), Galaxy Evolution: The Milky Way Perspective,
  ASP Conf. Ser. 49, Astronomical Society of the Pacific, San Francisco,  125

\bibitem[\protect\astroncite{Gilmore}{1999}]{gil99x}
Gilmore G. 1999, Baltic Astronomy 8, 203

\bibitem[\protect\astroncite{Gilmore et~al.}{2000}]{gbf+00}
Gilmore G., de~Boer K.~S., Favata F. et~al. 2000,
\newblock in J.~B. Brekinridge, P. Jakobsen (eds.), UV, Optical and IR Space
  Telescopes and Instruments, SPIE 4013

\bibitem[\protect\astroncite{Gilmore \& H{\o}g}{1995}]{gh95}
Gilmore G., H{\o}g E. 1995,
\newblock in M.~A.~C. Perryman, F. van Leeuwen (eds.), Future Possibilities for
  Astrometry in Space, ESA SP-379, ESA, Noordwijk, p.~95

\bibitem[\protect\astroncite{Gilmore et~al.}{1989}]{gwk89}
Gilmore G., Wyse R. F.~G., Kuijken K. 1989, ARA\&A 27, 555

\bibitem[\protect\astroncite{Hernandez et~al.}{2000}]{hgv99}
Hernandez X., Gilmore G., Valls-Gabaud D. 2000, MNRAS 316, 605

\bibitem[\protect\astroncite{H{\o}g}{1993}]{hog93}
H{\o}g E. 1993,
\newblock in I.~I. Mueller, B. Kolaczek (eds.), Developments in Astrometry and
  their Impact on Astrophysics and Geodynamics, IAU Symp.~156, Kluwer, The
  Netherlands, p.~37

\bibitem[\protect\astroncite{H{\o}g}{1995a}]{hog95a}
H{\o}g E. 1995a,
\newblock in E. H{\o}g, P.~K. Seidelmann (eds.), Astronomical and Astrophysical
  Objectives of Sub-Milliarcsec Astrometry, IAU Symp.~166, Kluwer,  317

\bibitem[\protect\astroncite{H{\o}g}{1995b}]{hog95b}
H{\o}g E. 1995b,
\newblock in M.~A.~C. Perryman, F. van Leeuwen (eds.), Future Possibilities for
  Astrometry in Space, ESA SP-379, ESA, Noordwijk,  263

\bibitem[\protect\astroncite{H{\o}g et~al.}{1999a}]{hfk+99}
H{\o}g E., Fabricius C., Knude J., Makarov V.~V. 1999a, Baltic Astronomy 8, 25

\bibitem[\protect\astroncite{H{\o}g et~al.}{1999b}]{hfm99}
H{\o}g E., Fabricius C., Makarov V.~V. 1999b, Baltic Astronomy 8, 233

\bibitem[\protect\astroncite{H{\o}g \& Lindegren}{1993}]{hl93}
H{\o}g E., Lindegren L. 1993,
\newblock in I.~I. Mueller, B. Kolaczek (eds.), Developments in Astrometry and
  their Impact on Astrophysics and Geodynamics, IAU Symp.~156, Kluwer, The
  Netherlands, ~31

\bibitem[\protect\astroncite{H{\o}g \& Lindegren}{1994}]{hl94}
H{\o}g E., Lindegren L. 1994,
\newblock in L.~V. Morrison, G. Gilmore (eds.), Galactic and Solar System
  Optical Astrometry, Cambridge University Press, p.~246

\bibitem[\protect\astroncite{Ibata et~al.}{1997}]{iwg+97}
Ibata R.~A., Wyse R. F.~G., Gilmore G., Irwin M., Suntzeff N. 1997, AJ 113, 634

\bibitem[\protect\astroncite{Irwin}{1985}]{irw85}
Irwin M.~J. 1985, MNRAS 214, 575

\bibitem[\protect\astroncite{{Johnston} \& {de Vegt}}{1999}]{kv99}
{Johnston} K.~J., {de Vegt} C. 1999, ARA\&A 37, 97

\bibitem[\protect\astroncite{Lattanzi et~al.}{2000}]{lss+00}
Lattanzi M.~G., Spagna A., Sozzetti A., Casertano S. 2000, MNRAS 317, 211

\bibitem[\protect\astroncite{Lebreton}{2000}]{leb00}
Lebreton Y. 2000, ARA\&A 38, in press

\bibitem[\protect\astroncite{Lindegren et~al.}{1993a}]{lbg+93a}
Lindegren L., Bastian U., Gilmore G. et~al. 1993a,
\newblock Roemer: Proposal for the Third Medium Size ESA Mission (M3),
\newblock Technical Report, Lund Observatory

\bibitem[\protect\astroncite{Lindegren \& Perryman}{1996}]{lp96}
Lindegren L., Perryman M. A.~C. 1996, A\&AS 116, 579

\bibitem[\protect\astroncite{Lindegren et~al.}{1993b}]{lpb+93b}
Lindegren L., Perryman M. A.~C., Bastian U. et~al. 1993b,
\newblock GAIA --- Proposal for a Cornerstone Mission concept submitted to ESA
  in October 1993,
\newblock Technical Report, Lund

\bibitem[\protect\astroncite{Majewski}{1993}]{maj93}
Majewski S. 1993, ARA\&A 31, 575

\bibitem[\protect\astroncite{Marcy \& Butler}{1998}]{mb98b}
Marcy G.~W., Butler R.~P. 1998,
\newblock in R.~A. Donahue, J.~A. Bookbinder (eds.), Cool Stars, Stellar
  Systems and the Sun; Proceedings of the 10th Cambridge Workshop, ASP Conf.
  Series 154, San Francisco, p.~9

\bibitem[\protect\astroncite{M\'erat et~al.}{1999}]{msc+99}
M\'erat P., Safa F., Camus J.~P., Pace O., Perryman M. A.~C. 1999, Baltic
  Astronomy 8, 1

\bibitem[\protect\astroncite{Mignard}{1999}]{mig99}
Mignard F. 1999,
\newblock in D. Egret, A. Heck (eds.), Harmonizing Cosmic Distance Scales in a
  Post-Hipparcos Era, ASP Conf. Ser. 167, p.~44

\bibitem[\protect\astroncite{Munari}{1999a}]{mun99b}
Munari U. 1999a, Baltic Astronomy 8, 73

\bibitem[\protect\astroncite{Munari}{1999b}]{mun99a}
Munari U. 1999b, Baltic Astronomy 8, 123

\bibitem[\protect\astroncite{O'Mullane \& Lindegren}{1999}]{ml99}
O'Mullane W., Lindegren L. 1999, Baltic Astronomy 8, 57

\bibitem[\protect\astroncite{Paczy\'nski}{1997}]{pac97}
Paczy\'nski B. 1997,
\newblock in R. Ferlet, J.~P. Maillard, B. Raban (eds.), Proc. 12th IAP
  Astrophysics Coll. 1996: Variable Stars and Astrophysical Returns from
  Microlensing Surveys, Editions Fronti\`eres,  357

\bibitem[\protect\astroncite{Perryman}{2000}]{per00}
Perryman M. A.~C. 2000, Rep. Prog. Phys. 63, 1209

\bibitem[\protect\astroncite{R{\"o}ser et~al.}{1997}]{rbb+97}
R{\"o}ser S., Bastian U., de~Boer K.~S. et~al. 1997,
\newblock in M.~A.~C. Perryman, P.~L. Bernacca (eds.), Hipparcos Venice 97, ESA
  SP-402, ESA, Noordwijk, p.~777

\bibitem[\protect\astroncite{S\"oderhjelm}{1999}]{sod99}
S\"oderhjelm S. 1999, A\&A 341, 121

\bibitem[\protect\astroncite{Straizys}{1999}]{str99}
Straizys V. 1999, Baltic Astronomy 8, 109

\bibitem[\protect\astroncite{{Triebes} et~al.}{2000}]{tgh+00}
{Triebes} K.~J., {Gilliam} L., {Hilby} T., {Horner} S.~D., {Perkins} P.,
  {Vassar} R.~H., {Harris} F.~H., {Monet} D.~G. 2000, SPIE 4013  482--492

\bibitem[\protect\astroncite{Vaccari}{2000}]{vac00}
Vaccari M. 2000,
\newblock GAIA Galaxy Survey, Master Thesis,
\newblock University of Padova

\bibitem[\protect\astroncite{Wyse et~al.}{1997}]{wgf97}
Wyse R. F.~G., Gilmore G., Franx M. 1997, ARA\&A 35, 637

\end{thebibliography}

\end{document}